\documentclass[12pt,a4paper]{article}

\usepackage[english]{babel}
\usepackage[utf8x]{inputenc}
\usepackage[T1]{fontenc}
\usepackage[a4paper,top=3cm,bottom=2cm,left=3cm,right=3cm,marginparwidth=1.75cm]{geometry}

\usepackage{amsmath}
\usepackage{graphicx}
\usepackage{amssymb}
\usepackage{hyperref}
\usepackage{authblk}
\usepackage{caption}
\usepackage{tabularx}
\usepackage{cite}

\graphicspath{{Pics/}}

\newcommand{\be}{\begin{equation}}
\newcommand{\ee}{\end{equation}}
\newcommand{\bea}{\begin{eqnarray}}
\newcommand{\eea}{\end{eqnarray}}
\newcommand{\lc}{\left[}
\newcommand{\rc}{\right]}
\newcommand{\lp}{\left(}
\newcommand{\rp}{\right)}

\newcommand{\MM}{\left| \mathcal{M} \right|^2}
\newcommand{\dNeff}{\Delta N_{\rm eff}}
\newcommand{\Ht}{\tilde{H}}
\newcommand{\gt}{\tilde{g}}
\newcommand{\ft}{\tilde{f}}
\newcommand{\bGamma}{\bar{\Gamma}} 

\title{Precise predictions for the QCD axion contribution to dark radiation with full phase-space evolution}
\author[]{Marcin Badziak}
\author[]{Maxim Laletin}
\affil[]{Institute of Theoretical Physics, Faculty of Physics, University of Warsaw, ul. Pasteura 5, PL-02-093 Warsaw, Poland}
\date{}

\begin{document}
\maketitle

\begin{abstract}
We compute the QCD axion contribution to the energy density of dark radiation, parameterized by $\dNeff$, by solving Boltzmann equations for the momentum distribution functions including the effects of quantum statistics for all particles involved in the axion production processes. This approach gives precise prediction for $\dNeff$ independently of whether axions are produced via freeze-out or freeze-in. We focus on axions produced via flavor-conserving and flavor-violating interactions with leptons. Our precise predictions for $\dNeff$ can differ from those assuming thermal shape for the momentum distribution functions, as commonly done in the literature, by more than the experimental precision of future Cosmic Microwave Background (CMB) observations. Current lower limits on the axion couplings from Planck constraints on $\dNeff$  are also affected by our precise computation which, in particular, results in a strongly relaxed bound on flavor-violating axion couplings to tau lepton and muon or electron.  

\end{abstract}

\section{Introduction}

The QCD axion is a light pseudoscalar particle~\cite{Weinberg:1977ma,Wilczek:1977pj} predicted by a Peccei-Quinn (PQ) solution~\cite{Peccei:1977hh,Peccei:1977ur} to the strong CP problem and is a good candidate for cold dark matter (DM) candidate~\cite{Preskill:1982cy,Abbott:1982af,Dine:1982ah}. 

The strength of the axion interactions to the Standard Model (SM) particles is controlled by the axion decay constant $f$ which also fixes the small axion mass $m_a$
generated by non-perturbative QCD effects: $m_a \approx 6~\text{eV} \, (f/10^6~\rm GeV)$~\cite{GrillidiCortona:2015jxo}.
The axion decay constant is constrained by astrophysics and the lower bound in typical QCD axion models, such as KSVZ~\cite{Kim:1979if,Shifman:1979if} or DFSZ~\cite{Dine:1981rt,Zhitnitsky:1980tq}, is $\mathcal{O}(10^8)$~GeV~\cite{Carenza:2019pxu,Buschmann:2021juv} but one should note that these bounds rely on effects that cannot be calculated from first principles so the uncertainties are large and constraints obtained by different groups differ between each other, cf.~e.g.~\cite{Chang:2018rso,Bar:2019ifz,Carenza:2020cis,Carenza:2019pxu,Buschmann:2021juv,Springmann:2024mjp}. Even taking the astrophysical constraints at face value in the so-called astrophobic axion models $f$ may be as small $\mathcal{O}(10^6)$~GeV~\cite{DiLuzio:2017ogq,Bjorkeroth:2019jtx,Badziak:2021apn,Badziak:2023fsc,Badziak:2024szg}. 
Thus, the QCD axion mass cannot be larger than few eV and typically is even smaller.
In spite of its small mass the axion is a good candidate for cold DM because it is generically produced non-thermally.  

QCD axions can be also produced thermally and
such axions are relativistic at least at the early stage of the evolution of the Universe~\cite{Chang:1993gm}. For small enough $f$ axions are brought to thermal equilibrium with the SM plasma and their abundance freezes out when they decouple. Thermal axions can be constrained by the Big Bang Nucleosynthesis (BBN) and cosmological observations, see~e.g.~\cite{Hannestad:2005df,DEramo:2022nvb,Yeh:2022heq}, which are more robust than astrophysical constraints. Since they are typically relativistic they contribute to the total energy density of relativistic degrees of freedom, hence playing the role of dark radiation, which is conveniently parameterized  by the effective number of neutrino species $N_{\rm eff}$. The new physics contribution to $N_{\rm eff}$, typically denoted by $\dNeff$,  is constrained by the observations of the Cosmic Microwave Background (CMB) and baryon acoustic oscillations (BAO). The most recent combined analysis by Planck collaboration sets the upper bound $\dNeff\leqslant0.3$ at 95\% confidence level (CL)~\cite{Planck:2018vyg}. This bound is expected to be improved to $0.1$ at Simons Observatory \cite{SimonsObservatory:2018koc} and eventually to $0.05$ by CMB-S4 experiments \cite{CMB-S4:2016ple}.

This tremendous progress on observational side can be fully exploited to constrain axions only in the presence of precise theoretical predictions for the energy density of thermally produced axions. It has been recognized that precise determination of $\dNeff$ requires solving Boltzmann equations to account for a change in the number of relativistic degrees of freedom during axion decoupling, see e.g. Refs.~\cite{Ferreira:2018vjj,DEramo:2018vss,Arias-Aragon:2020shv,Ferreira:2020bpb,Green:2021hjh,DEramo:2021usm,DEramo:2021psx,DEramo:2021lgb,Badziak:2024szg}. However, in the majority of literature it has been assumed that momentum distribution for axions follows thermal Maxwell-Boltzmann (MB) one. Such an assumption greatly simplifies numerical computations because it is then enough to solve momentum-independent Boltzmann equations for the number density and use an analytical relation between the energy density and number density of axions. However, such a simplified approach does not take into account several effects that make the axion momentum distribution different than the thermal one and may impact the final energy density of axions. In particular, such simplified approach is not justified when axions are not thermalized but rather produced via freeze-in~\cite{Hall:2009bx} or when the number of entropy degrees of freedom rapidly changes during axion decoupling. Indeed, it was recently demonstrated for the case of axion scattering with pions that solving momentum-dependent Boltzmann equations for the distribution functions leads to a prediction of $\dNeff$ that differs from that in the simplified approach by more than the expected sensitivity of CMB-S4~\cite{Notari:2022ffe}. The more precise computation allowed Ref.~\cite{Notari:2022ffe} to improve the hot dark matter bound on the KSVZ model, see also Ref.~\cite{Bianchini:2023ubu} where the bound was updated with the latest cosmological data sets.\footnote{Momentum-dependent Boltzmann equations have recently been used to predict $\dNeff$ from axions also in other papers.
For example, Ref.~\cite{Dunsky:2022uoq} used it to find dark radiation constraints for heavy QCD axion while Ref.\cite{Bouzoud:2024bom} used it to compute $\dNeff$ from  axion-gluon interactions.} However, those results are not enough to set constraints on axion models in which other channels for axion thermal production are non-negligible, especially those in which axion-pion coupling is suppressed as it is the case, for example, in astrophobic axion models.

The goal of the present article is to compute precisely the contribution to $\dNeff$ from axions produced in other relevant production channels using full momentum-dependent solution to Boltzmann equations for the distribution functions. We perform this computation in a model-independent way for axions produced via most relevant interactions with leptons. For flavor-conserving interactions, we consider  scattering with muons or taus which are important production processes, for example, in DFSZ models~\cite{Ferreira:2020bpb}.  In addition we perform the computation for flavor-violating $\tau$ decays to axion and electron or muon. Flavor-violating axion couplings are characteristic for models with flavor non-universal PQ charge assignment such as those explaining fermion mass hierarchies~\cite{Ema:2016ops,Calibbi:2016hwq}, or minimal astrophobic axion models~\cite{DiLuzio:2017ogq,Badziak:2024szg}.     We do not consider axion production via scattering with electrons or flavor-violating $\mu\to ea$ decays due to strong laboratory constraints on the axion-electron couplings from XENONnT~\cite{XENON:2022ltv}, LZ~\cite{LZ:2023poo} and PandaX-4T~\cite{PandaX:2024cic} that searched for axions produced in the Sun and the axion-electron-muon coupling from TRIUMF~\cite{Jodidio:1986mz} and TWIST~\cite{TWIST:2014ymv} (see~Ref.~\cite{Calibbi:2020jvd} for the overview of the experimental constraints on the flavor-violating axion couplings). We focus on axion-lepton interactions because they do not suffer from uncertainties due to unknown non-perturbative effects that spoil the precision of $\dNeff$ computation for axion production processes involving quarks, see Ref.~\cite{Notari:2022ffe} for a recent discussion on this issue. 

We use our improved results to set model-independent constraints on axion couplings to muons and the flavor-violating ones to $\tau-e$ and $\tau-\mu$. We also estimate sensitivity of future CMB experiments for these couplings and show that CMB-S4 will be able to constrain also axion coupling to taus.  
 
The rest of the article is organized as follows. In Section~\ref{Sec:Evolution} we introduce the two approaches to calculate the relic density of axions and their energy distribution that we compare in our work. In Section~\ref{Sec:Processes} we discuss the types of axion production processes of interest and provide the details of their implementation. In Section~\ref{Sec:shape_and_dneff} we discuss how the $\dNeff$ is calculated and analyze the impact of the axion distribution function. In Section~\ref{Sec:Results} we present and compare the results of our computations for the two methods on different axion processes. We conclude our work and outline the prospects for further research in Section~\ref{Sec:Conclusion}. In Appendices~\ref{apx:decay} and \ref{apx:2to2} we derive the expressions for the collision terms of 2-body decays and $2 \rightarrow 2$ processes with an axion, respectively. 
  
\section{Axion production and evolution in the early Universe}
\label{Sec:Evolution}

We are going to consider two approaches to calculating the properties of axion population and their evolution in the early Universe.

\subsection{The Boltzmann equation for the number density (nBE)}
\label{Sec:nBE}

The Boltzmann equation for the number density (nBE) is one of the simplest approaches to calculate the density of axions (and many other particles in general) in the early Universe. This method is well elaborated in the classical work of Gelmini \& Gondolo \cite{Gondolo:1990dk} and has long become a standard way of describing the thermal history of BSM species. The nBE describes the evolution of the density of $a$ particles in time $t$ in the expanding Friedmann–Lemaître–Robertson–Walker (FLRW) Universe characterized by the Hubble parameter $H(t)$

\be
\frac{dn_a}{dt} + 3H n_a = \bar{\Gamma}^+_a - \bGamma^-_a \, ,
\label{eq:classic_nbe}
\ee
where $\bGamma^+_a$ and $\bGamma^-_a$ are the rates\footnote{We use the notion $\bGamma$ for rates (GeV$^4$ in natural units) to discriminate them from the decay width $\Gamma$ (GeV in natural units).} of production and destruction of $a$ particles in the hot plasma. The rates can be functions of $t$ and $n_a$ (especially the destruction rate). Eq.~\eqref{eq:classic_nbe} can be conveniently rewritten in terms of the (inverse) temperature variable $x = m/T$ and the comoving density $Y_a = n_a/s$, where $s$ is the entropy density of the Universe and $m$ can be the mass of any particle, but it is the most convenient to use the mass parameter that sets the characteristic scale for the main reaction under consideration\footnote{In our analysis $x$ is defined via the mass of the heaviest particle in the reaction(s) that we study.}  

\be
sx\Ht \, \frac{dY_a}{dx} = \bGamma^+_a - \bGamma^-_a \,
\label{eq:nBE_gen}
\ee
where $\Ht = H/(1+\gt)$ is the Hubble parameter divided by 

\be
1 + \gt = 1 - \frac{1}{3} \frac{d\ln h_s}{d\ln x} \, 
\ee
with $h_s$ being the number of entropy degrees of freedom\footnote{In this work we used a spline parametrization of $h_s$ taken from Ref.~\cite{Borsanyi:2016ksw}. As was pointed out in Refs.~\cite{Badziak:2024szg,DEramo:2018vss} various parametrizations have a tiny effect on the solution of nBE, although it might have a larger impact on the solution of fBE.}. The left-hand side of the nBE essentially describes the cosmological impact on the population of $a$ particles, while the right-hand side describes the intensity of particle physics (number-changing) processes involving $a$ particle. The comoving density $Y_a$ is conserved when all these processes effectively seize. 

As a simple example that is relevant for our study let us consider the annihilation of two leptons $l$ into an axion $a$ and a photon $\gamma$. The production rate for this process at a given $x$ depends mainly on the densities of $l^+$ and $l^-$ and on the intensity of annihilation that can be characterized by the velocity-averaged cross-section $\langle \sigma_{\rm ann} v\rangle$

\be
\bGamma^+ = n_{l^+}n_{l^-} \langle \sigma_{\rm ann} v\rangle \, .
\ee
The corresponding destruction process is the inverse annihilation of axion and photon into a pair of leptons and the rate of destruction is proportional to the (unknown) density of $a$ particles

\be
\bGamma^- = n_{a}n_{\gamma} \langle \sigma_{\rm inv} v\rangle \, .
\ee
If the system is close to thermal equilibrium\footnote{Precisely, if the axion distribution has the thermal shape. See Appendix~\ref{apx:2to2} for more details.} the velocity-averaged cross-section for forward and inverse annihilation is essentially the same and the right-hand side of the nBE can be written as 

\be
\bGamma^+_a - \bGamma^-_a = n_{l^+}n_{l^-} \langle \sigma_{\rm ann} v\rangle \left( 1 - \frac{Y_a}{Y^{\rm eq}_a} \right) \, ,
\ee
where $Y^{\rm eq}_a$ is the equilibrium comoving abundance of axions at the given $x$ and the expression in the brackets follows from the fact that the rates $\bGamma^+_a$ and $\bGamma^-_a$ are equal in the equilibrium. 

From the computational point of view, nBE is an ordinary differential equation (ODE) in $x$, which can be rather simply solved using numerous computational tools and mathematical frameworks and is implemented in several common astroparticle physics packages (which can also handle complicated models with several particles and various interactions between them that leads to a system of nBEs) such as micrOMEGAs \cite{Alguero:2023zol}, MadDM \cite{Ambrogi:2018jqj} or DarkSUSY \cite{Bringmann:2018lay}. However, the lucidity of the nBE approach can be misleading as it covers several assumptions. 

The most important of them is the assumption that the $a$ particles are in \textit{kinetic equilibrium} with the surrounding plasma, i.e. the \textit{shape} of their energy distribution is the same as of the corresponding equilibrium distribution with the normalization factor being $Y_a/Y^{\rm eq}_a$. The kinetic equilibrium is typically maintained via elastic scatterings of $a$ particles on the particles in the SM plasma, which is justified in case of many BSM models, however not necessarily in the case of axions. For example, most WIMP models contain vertices with two WIMPs and a mediator, so the cross section of annihilation to SM ($s$-channel) and elastic scattering ($t$-channel) processes at tree level have the same order of the coupling associated with that vertex. Due to the shift symmetry the axion vertices with SM particles can contain only a single axion branch, hence the cross section of the elastic scatterings is suppressed by a factor of $f^{-2}$ w.r.t. to the production processes.
Since lower bound on $f$ is at least $10^6$~GeV elastic scatterings may not be able to bring axions into kinetic equilibrium.
The shape of the distribution in some cases can drastically affect the rates, hence requiring a more general approach (see Refs.~\cite{Binder:2017rgn,Hryczuk:2022gay} for examples). 

The second important approximation which is often exploited is to neglect the quantum corrections, which means that the plasma has no impact on the rates of these reactions. While it is generally true for classical, non-relativistic or very dilute gases, quantum corrections can affect the rates involving relativistic particles and lead to a deviation of the predicted density. 
The assumption of vanishing quantum corrections is coherent with using the classical distribution function, i.e. the MB distribution, for the particles in the reactions. Altogether these approximations drastically simplify the structure of the right-hand side of the nBE and the computation of the rates. The impact of quantum corrections on the abundance of dark radiation, and axions in particular, has been shown to be non-negligible~\cite{Olechowski:2018xxg,Green:2021hjh,DEramo:2023nzt}.

It is also important to stress that the outcome of the solution to nBE is only the number density of axions while cosmologically relevant quantity is the energy density of axions parameterized by $\dNeff$. Calculation of $\dNeff$ in this simplified approach relies on the assumption that the energy density of axions is given by the equilibrium distribution which leads to a simple relation between $\dNeff$ and $Y_a$. However, this assumption is not valid in general. In particular, such a simplified approach leads to an underestimation of $\dNeff$ when axions are produced via freeze-in~\cite{Hall:2009bx}, as recently emphasized in Ref.~\cite{Badziak:2024szg}. To go beyond the limitations of the standard nBE approach one needs to consider a more general Boltzmann equation for the full phase-space distribution function which we discuss in the following subsection\footnote{Other intermediate approaches are possible, e.g. see the coupled Boltzmann equation approach in Refs.~\cite{Binder:2017rgn,Binder:2021bmg}.}.     

\subsection{The full Boltzmann equation (fBE)}
\label{Sec:fBE}

The general Boltzmann equation for a particle $a$ describes the evolution of the distribution function in the phase space $f_a$ in the expanding FLRW Universe. In the comoving coordinates $x = m/T$ and $q = p/T$ it has the following form 

\be
\Ht (x\partial_x - \gt q\partial_q)f_a(x,q) = C[f_a] \, ,
\label{eq:FullBoltzmann}
\ee
where $C[f_a]$ is the collision term that takes into account the impact of all the particle physics processes in which particle $a$ participates on its momentum distribution. 
The impact of the cosmological evolution on the distribution function is contained in the left-hand side of the full Boltzmann equation. Note that unlike the case of nBE the rate of change of entropy degrees of freedom $\gt$ is manifested more than just in $H/(1 + \gt)$ and can have a bigger effect on the solution. The collision term is the integral over the phase space of all the particles that participate in a reaction that involves a particle $a$ (limited by the 4-momentum conservation and the physics of interaction, manifested in the amplitude-squared $\MM$) excluding the phase space of particle $a$ itself. For processes that involve two or more $a$ particles the collision term contains at least one unknown function $f_a$ under the phase-space integral, thus in general fBE is an integro-differential equation.

Integrating the left-hand side of the fBE over the momentum-space of the $a$ particle directly leads to the left-hand side of the nBE in Eq.~\eqref{eq:nBE_gen}. Integrating the collision term in the same way is generally a very intricate process, especially since the collision term can contain an unkown distribution function $f_a$. Such an integration is only straightforward for the processes that do not change the number of $a$ particles (like elastic scatterings) as they have no impact on the density of $a$ gas and hence can be integrated out. As the axion processes have only one axion vertex at the first order in $(1/f)^2$, one can show that the structure of the collision term is the following (see Appendices ~\ref{apx:decay} and ~\ref{apx:2to2})

\be
C[f_a] = \frac{1}{2g_a E_a} \left( 1 - \frac{f_a}{f^{\rm eq}_a} \right)\gamma(x,q) \, ,
\label{eq:collterm_general_axion}
\ee
where $g_a$ is the axion number of degrees of freedom\footnote{Here and below in the text we use the general expressions that include the axion number of degrees of freedom, even though the models that we consider contain only one scalar axion, therefore $g_a = 1$.}, $E_a$ is the axion energy and $\gamma$ can be roughly regarded as a differential rate of axion production per unit of axion phase-space in a given reaction. Hence, for axions the fBE is a partial differential equation (PDE) in $x$ and $q$. If the number of entropy degrees of freedom is not changing Eq.~\eqref{eq:FullBoltzmann} for axions simplifies to an ODE in $x$. For the purpose of numerical implementation of the fBE it is convenient to rewrite Eq.~\eqref{eq:FullBoltzmann} in terms of the rescaled distribution function $\ft_a(x,q) = q^2 \cdot f_a(x,q)$ as follows

\be
\Ht \lc x \partial_x - \gt \lp q \partial_q - 2 \rp \rc \ft_a(x,q) = q^2 C \, [\ft_a] \, .
\label{eq:FBEredef}
\ee
The immediate utility of this definition can be seen from the simplicity of the boundary condition: 
$\ft_a(x,0) = \ft_a(x,\infty) = 0$. Other useful features include the fact that $\ft_a$ is better for the purpose of comparing the actual solution with the equilibrium-shaped distribution and that $\ft_a$ can be plainly integrated over $q$ to obtain the density of axions

\be
n_a = g_a \int \frac{d^3p_a}{(2\pi)^3} \, f_a = \frac{g_a T^3}{2\pi^2} \int dq \; \ft_a \, .
\ee

In the absence of any interactions (for free particles) the solution of Eq.~\eqref{eq:FullBoltzmann} can be simply expressed as

\be
f^{\rm free}_a(x,q) = f_a^{(0)}\left( q \cdot \left[ \frac{h_s(x_0)}{h_s(x)}\right]^{1/3}\right) ,
\label{eq:free_particle_PDF}
\ee
where $x_0$ is the initial value of $x$ (or the moment of decoupling of $a$ particles from the interactions, for example) and $f^{(0)}_a$ is the distribution function that correspond to $x_0$. In other words, in the absence of interactions the temperature of the gas of $a$ particles decreases w.r.t. to the temperature of the plasma as the entropy of the $a$ particle population is diluted by the particles that go out of equilibrium, so the distribution function just shifts to the region of smaller momenta in the $q$-space. The same phenomenon occurs with neutrinos in the standard cosmology (see e.g. Ref.~\cite{Lesgourgues:2006nd}).

The existing publicly available numerical fBE solvers for DM are quite scarce with the most prominent example being the DRAKE code \cite{Binder:2021bmg}, which in its current implementation is limited to WIMPs. We refer the reader to the Appendices~\ref{apx:decay} and~\ref{apx:2to2} for more details regarding the structure of the collision term and to Ref.~\cite{Binder:2017rgn} for a more comprehensive description of the fBE approach. 

\section{Axion production processes}
\label{Sec:Processes}

We focus on the axions  produced via interactions with leptons. The relevant part of the effective Lagrangian that describes axion couplings well below the PQ breaking scale reads
\begin{equation}
{\cal L}  \supset \frac{\partial_\mu a}{2 f} \, \overline{l}_i \gamma^\mu \left( C^V_{ij} + C^A_{ij} \gamma_5 \right) l_j \, ,
\end{equation}
where $f$ is the axion decay constant.\footnote{We denote the axion decay constant as $f$ skipping subscript 'a' to avoid confusion with the axion distribution function $f_a(x,q)$.} For flavor-violating couplings we define $C_{ij} \equiv \sqrt{|C_{ij}^A|^2 + |C_{ij}^V|^2}$ since thermal axion production typically does not depend on the chirality structure of axion couplings. For flavor-diagonal couplings we define $C_i\equiv C_{ii}^A$.

We first discuss flavor-violating $\tau$ decays since for $\mathcal{O}(1)$ couplings they  produce axions more efficiently than flavor-conserving lepton scatterings which we discuss subsequently. 

\subsection{Flavor-violating interactions}

Flavor-violating axion interactions appear in models in which the PQ charge assignment is non-universal for fermion generations. Such models have been proposed to explain the SM fermion mass hierarchies by associating the PQ symmetry with the flavor symmetry~\cite{Ema:2016ops,Calibbi:2016hwq}. Flavor non-universal PQ charge assignment is also a key feature of astrophobic axion models~\cite{DiLuzio:2017ogq,Bjorkeroth:2019jtx,Badziak:2021apn,Badziak:2023fsc,Badziak:2024szg}. In particular, suppression of axion couplings to nucleons and electrons in the minimal astrophobic axion models proposed in Ref.~\cite{DiLuzio:2017ogq} requires $C_{\tau e}$ to be $\mathcal{O}(1)$~\cite{Badziak:2021apn,Badziak:2024szg}. 

Flavor non-diagonal interactions of leptons with axion lead to the decays of heavier leptons into an axion and a lighter lepton. We consider tau decays into a muon and an axion

\be
\tau^{\pm} \rightarrow \mu^{\pm} a \, .
\ee
The analysis of $\tau \rightarrow e a$ decay is similar since both muon and electron mass is negligible as compared to the tau mass. We do not consider $\mu\to e a$ decays due to extremely strong lower bounds on the muon-electron-axion coupling from colliders. 
In general non-diagonal interactions also lead to inelastic scatterings
but the contribution from these processes to the axion abundance is suppressed w.r.t. decays for temperatures for which the axion production mainly occurs i.e. slightly below the tau mass~\cite{Aghaie:2024jkj}.\footnote{Note that a calculation of the flavor-violating scattering requires a proper treatment of infrared divergences in the limit of massless lighter lepton (muon in this case) which otherwise leads to unphysical enhancement of the scattering rate. A naive regularization of these divergencies by using thermal mass, as it was done in Ref.~\cite{DEramo:2021usm}, leads to a large overestimate of the thermal axion production via flavor-violating scattering~\cite{Aghaie:2024jkj}. } 

The right-hand side of the nBE for decays is 

\be
\bGamma^+_a - \bGamma^-_a = 2\Gamma_{\tau \rightarrow \mu a} \frac{K_1(x)}{K_2(x)}  s Y_\tau \Bigg( 1 - \frac{Y_a}{Y^{\rm eq}} \Bigg) \, ,
\label{eq:rhs_nbe_decay}
\ee
where $x = m_{\tau}/T$, $K_1$ and $K_2$ are the modified Bessel functions of the first and the second kind respectively, $Y_\tau$ is the comoving density of tau leptons and the factor of 2 in front of the expression takes into account the contribution to axion density from both $\tau$ particles and antiparticles. 
$\Gamma_{\tau \rightarrow \mu a}$ is the partial width of the corresponding $\tau$ decay channel (averaged over the initial and final spin states) 

\be
\Gamma_{\tau \rightarrow \mu a} = \lp \frac{C_{\tau\mu}}{f} \rp^2 \frac{m^3_{\tau}}{64\pi} \lp 1 - r^2 \rp^3 \, , 
\ee
where $r = m_{\mu}/m_{\tau}$ is the ratio of the product particle (muon) mass and the mother particle (tau) mass.  

The collision term for the decay can be calculated analytically without any assumptions. The result is given by Eq.~\eqref{eq:collterm_general_axion} with $\gamma (x,q)$ being

\be
\gamma_{\rm dec} = 2 \cdot \frac{2 g_{\tau} m_{\tau} \Gamma_{\tau \rightarrow \mu a} }{q \lp 1 - r^2 \rp} \; f^{\rm eq}_a(q) \, \log \lc \frac{1 + \exp(-\xi_1)}{1 + \exp(-\xi_2)}\rc \, ,
\label{eq:gamma_dec_main}
\ee
where $\xi_1$ and $\xi_2$ are the functions of $x$ and $q$ that account for the kinematical limits in the decay and are given by the following expressions

\bea
\xi_1 &=& \max \left\{ x , \; x \cdot r + q , \; \frac{x^2}{4q} \lc 1 + \frac{4q^2}{x^2\lp1 - r^2\rp} - \mu^2 \rc \right\} \, , \label{eq:xi_dec_1} \\
\xi_2 &=& \max \left\{ x-q , \; x \cdot r , \; \frac{x^2}{4q} \right\} \, . \label{eq:xi_dec_2}
\eea

\subsection{Flavor-conserving interactions}

Flavor diagonal interactions of leptons with axion lead to two main processes: annihilation and Primakoff scattering. 

\bea
\mu^+ \mu^- & \rightarrow & a \, \gamma \, , \\
\mu^{\pm} \; \gamma & \rightarrow & \mu^{\pm} a \, .
\eea
In this subsection we focus on interactions with muons but the same methodology applies to interactions with tau leptons for which we also present the results in section~\ref{Sec:Results}. Axion production via diagonal interactions with electrons is negligible for values of $f$ consistent with experimental constraints so we do not analyse them. 

The rates of these processes for the nBE are given by 

\be
\bGamma^+_a - \bGamma^-_a = \bigg( \bGamma_{\rm ann}(x) + 2 \cdot \bGamma_{\rm Prim}(x) \bigg) \lp 1 - \frac{Y_a}{Y^{\rm eq}_a} \rp \, ,
\label{eq:rhs_nBE_diagonal}
\ee
where $\bGamma_{\rm ann} = n_{l}^2 \langle \sigma_{\rm ann} v\rangle$ is the averaged rate of annihilations, $\bGamma_{\rm Prim} = n_{l}n_{\gamma} \langle \sigma_{\rm Prim} v\rangle$ is the averaged rate of Primakoff scatterings and the factor of 2 takes into account the scatterings on both muons and antimuons. 

The collision term for a $i + j \rightarrow k + a $ processes does not have a general analytical form as in the case of decays and is given by Eq.~\eqref{eq:collterm_general_axion} with $\gamma (x,q)$ being a 4-dimensional integral

\be
\gamma_{ij \rightarrow ak} = \frac{1}{p_a} \int dE_k \; \frac{\lp 1 \pm f_k(E_k) \rp}{16 \, (2\pi)^4} \int \frac{ds}{p^*_k\sqrt{s}} \int dt \MM \int d\cos{\phi} \; \frac{f_i^* \cdot f_j^*}{\sqrt{1 - \cos{\phi}^2}} \, ,
\label{eq:gamma_2to2_gen}
\ee
where $E_k$ is the energy of the $k$ particle (in the plasma frame),  $s$ and $t$ are the Mandelstam variables, the star denotes the variables in the CM frame, $\MM$ is the squared amplitude of the process (generally, a function of $s$ and $t$), $\phi$ is the angle between the projections of $\vec{p}_a$ and $\vec{p_i^*}$ on the plane that is orthogonal to $\vec{p_a^*}$. 
$f_i^*$ and $f_j^*$ denote the distribution functions in which the corresponding energies are Lorentz-transformed to the CM frame (see Appendix~\ref{apx:2to2}).

If the distribution functions $f_i$ and $f_j$ can be well approximated by the MB distribution Eq.~\eqref{eq:gamma_2to2_gen} drastically simplifies to 

\be
\gamma_{ij \rightarrow ak} = \frac{\exp(-q)}{(2\pi)^2} \int dE_k \; E_k \, f_k(E_k) \int ds \; \sigma_{ak \rightarrow ij} (s) \, v_{\rm Mol} \, ,
\label{eq:gamma_2to2_simp}
\ee
where $\sigma_{ak \rightarrow ij}$ is the cross section of the process and $v_{\rm Mol}$ is the Moller velocity in Eq.~\eqref{eq:MollerVelocity}. For annihilations and Primakoff scatterings the integral over $s$ can be calculated analytically leading to a simple 1-dimensional numerical integration formula for $\gamma(x,q)$. 

\begin{figure}[t]
	\center{\includegraphics[width=1\textwidth]{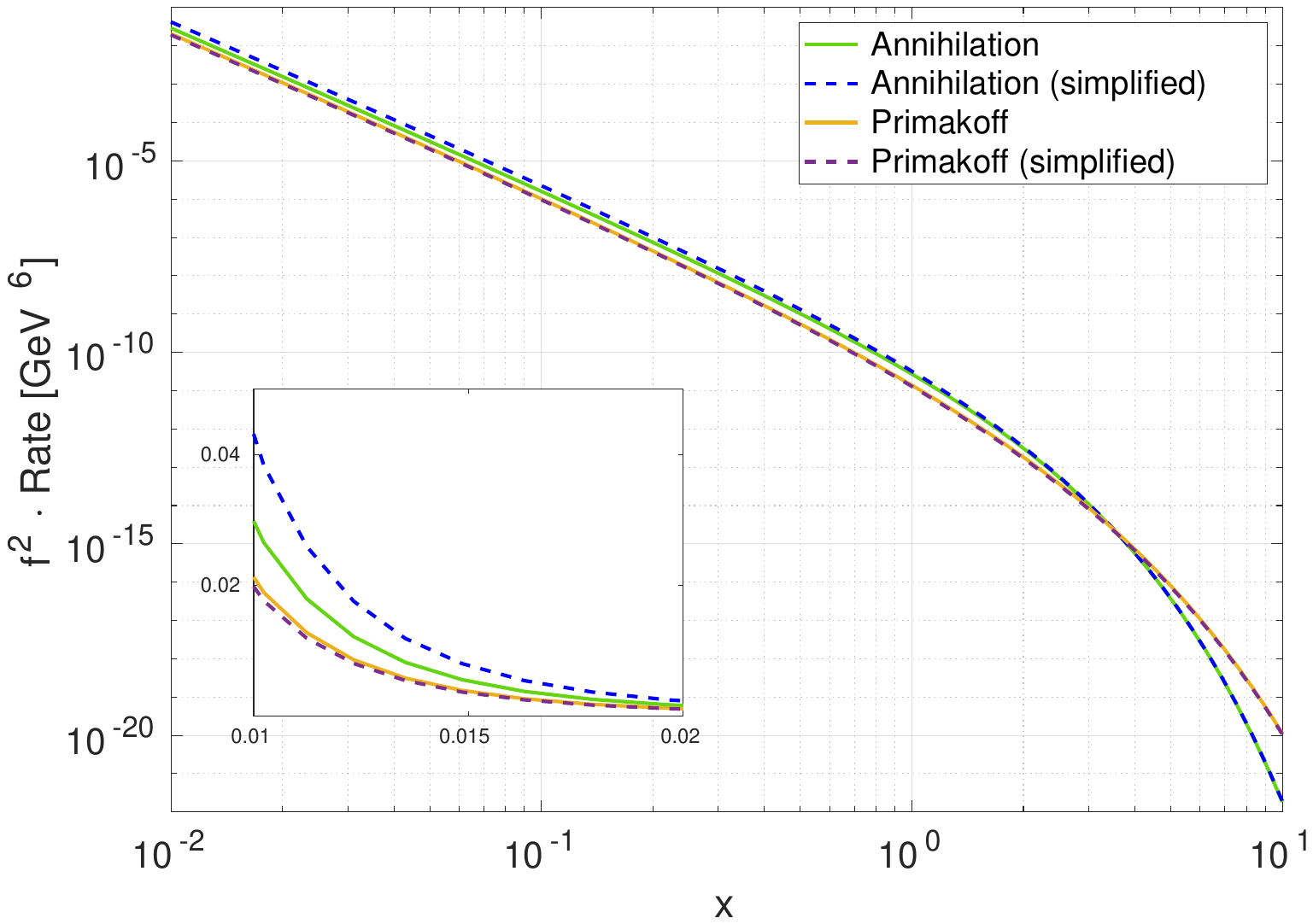}}
	\caption{The integral rate $f^2 \cdot \int d\Pi_a \, \gamma(x,q)$ for muon annihilation and Primakoff scattering as a function of x (independent of $f$). Solid lines correspond to the general Eq.~\eqref{eq:gamma_2to2_gen}, while the dashed ones to the simplified Eq.~\eqref{eq:gamma_2to2_simp}. The smaller plot in the bottom left corner zooms in on the upper left corner of the figure (linear scale). See the text for more details.}
	\label{fig:fa2_rate_muons}
\end{figure}

In Fig.~\ref{fig:fa2_rate_muons} we compare the two approaches to calculating the collision term. As $\gamma_{ij \rightarrow ak}$ is a function of $x$ and $q$ it is more convenient to compare the corresponding rates that are the integral of $\gamma_{ij \rightarrow ak}$ over the relativistic phase space of axions. Note that Eq.~\eqref{eq:gamma_2to2_simp} yields the same rates $\bGamma$ that are used in Eq.~\eqref{eq:rhs_nBE_diagonal} if the latter are calculated with the number densities that stem from the MB distribution functions. As one can see Eq.~\eqref{eq:gamma_2to2_simp} serves as a very good approximation especially for large values of $x$ and $q$. Despite the fact that the resulting rates computed in full generality do not deviate significantly from the simplified ones and predict essentially the same number densities of axions the distribution function given by the solution of fBE can still have a shape that is different from the thermal one and lead to a noticeable difference in $\dNeff$ (see Sec.~\ref{Sec:results_diagonal}).
 
Calculating the integrals in Eq.~\eqref{eq:gamma_2to2_gen} numerically is quite challenging and typically requires a lot of CPU time. For that purpose we developed a C++ code \texttt{CollCalc} \cite{CollCalc} that allows to compute these integrals efficiently\footnote{On average for a $1000 \times 100$ logarithmic space of relevant $x$ and $q$ values it takes about half an hour to compute on 8-core CPU.}. 

\section{$\dNeff$ and the shape of the axion distribution}
\label{Sec:shape_and_dneff}

Let us now discuss the computation of the axion contribution to the additional effective number of neutrino species $\dNeff$. We assume that axions are ultra-relativistic and their mass is negligible which is valid for temperatures much above the axion mass. We consider two approaches to compute $\dNeff$. 

The first approach assumes that axions are characterized by the equilibrium energy density at decoupling. In that case their energy density can be directly related to the comoving number density (which remains constant after decoupling) and $\dNeff$ in the late Universe is given by

\be
\dNeff = \frac{4}{7}\lp \frac{11}{4} \cdot \frac{2\pi^4 h_s(x)}{45 \zeta(3)} Y_a \rp^{4/3} \approx 74.85 \, Y^{4/3}_a \,.
\label{eq:dNeff_Simple}
\ee
This simplified approach is used in many works, see e.g. Refs.~\cite{Salvio:2013iaa,Ferreira:2018vjj,DEramo:2018vss,Arias-Aragon:2020shv,Ferreira:2020bpb,Green:2021hjh,DEramo:2021usm,DEramo:2021psx,Caloni:2022uya,Badziak:2024szg}. 
Eq.~\eqref{eq:dNeff_Simple} can be a good approximation for the axions generated via freeze-out process when the number of entropy degrees of freedom is not changing rapidly. However, it is not maintained in general, especially if the axions have never reached full thermal equilibrium but were rather produced via freeze-in. To demonstrate that let us start with a general formula for the axion contribution to the effective number of neutrino species

\be
\dNeff = \frac{8}{7} \lp \frac{11}{4} \rp^{4/3} \frac{\rho_a}{\rho_{\gamma}} \, .
\label{eq:dNeff_Exact}
\ee
Suppose axions have an equilibrium-shaped distribution with a normalization factor $A \equiv n_a / n^{\rm eq}_a$

\be
f_a = \frac{A}{\exp(E/T)-1} \, .
\ee
The corresponding number density and energy density can be calculated by integrating the axion distributions accordingly leading to

\be
n_a = A \cdot g_a \frac{\zeta(3)}{\pi^2} T^3 \, , \, \qquad \rho_a = A \cdot g_a \frac{\pi^2}{30} T^4 \, .
\ee
Now if we express the temperature of axions through their number density and substitute it in the expression for the density we get

\be
\rho_a = \frac{g_a}{A^{1/3}} \cdot \frac{\pi^2}{30} \left(\frac{\pi^2 n_a}{\zeta(3)g_a}\right)^{4/3} \, ,
\label{eq:axion_density_A}
\ee
which ultimately leads to a difference with the Eq.~\eqref{eq:dNeff_Simple} of a factor of $A^{1/3}$. Since for freeze-in axion production $A\ll1$ the simplified formula underestimates $\dNeff$ in that case. Furthemore, the shape of the axion distribution can deviate from the thermal shape, which leads to additional modifications of $\dNeff$ and require a more elaborated approach. 

The second and more accurate approach that we consider relies on the direct calculation of the axion energy density in the late Universe

\be
\rho_a = \frac{g_a}{2\pi^2} \int dE_a \, E_a^3  \, f_a(E_a) \,.
\ee
Plugging this density into Eq.~\eqref{eq:dNeff_Exact} we get $\dNeff$. In the limit when axions are ultrarelativistic in the late Universe their distribution function $f_a(\tilde{q})$ of the rescaled momentum $\tilde{q} = q \cdot (h_s(x_0)/h_s(x))^{1/3}$ remains the same since decoupling at $x_0$ (see Eq.~\eqref{eq:free_particle_PDF} and the text around), hence one gets

\be
\dNeff = \frac{8}{7} \left( \frac{11}{4} \right)^{4/3} \frac{15 g_a}{2\pi^4} \left( \frac{h_s(x_{0})}{h_s(x)} \right)^{-4/3} \int^{\infty}_0 d\tilde{q} \;  \tilde{q}^3 f_a(\tilde{q}) \; \Big|_{x_{0}} \, .
\label{eq:dNeff_rel_integrated}
\ee
We use the above formula with $h_s=43/11$ (corresponding to entropy degrees of freedom after electron-positron annihilation) to accurately compute axion contribution to $\dNeff$. In the integral above we use axion distribution function obtained by numerically solving fBE up to $x_0$ \footnote{In our calculations of the axion relic distribution function we used different values of $x_0 \geq 20$ for different processes and coupling sizes. In all of these cases the axion density was established long before $x_0$.}.

Equation \eqref{eq:dNeff_rel_integrated} can be reliably used as long as the axion mass is much smaller than temperature. This assumption is certainly valid during BBN. However, during the recombination it is only valid up to $\mathcal{O}(0.1)$~eV, or equivalently for $f$ down to about $10^7$~GeV.   
Thus, our results for $\dNeff$ can be used to reliably compare with the CMB constraints on $\dNeff$ for axion masses up to $\mathcal{O}(0.1)$~eV. For larger masses our limits should be treated as conservative since a non-vanishing value of axion mass increases the energy density of axions around recombination which makes the constraints stronger. Cosmological constraints that take into account the effect of the axion mass and non-thermal shape of axion distribution will be presented elsewhere.

\section{Results}
\label{Sec:Results}

Let us now discuss the abundance and energy density of axions using the simplified approach and our precise approach for each production channel separately. We compare our results for $\dNeff$ with the current constraint from Planck which excludes $\dNeff$ below 0.3 at 95\% CL~\cite{Planck:2018vyg} and future expected limits from Simons Observatory~\cite{SimonsObservatory:2018koc} and CMB-S4~\cite{CMB-S4:2016ple} corresponding to about 0.1 and 0.05, respectively, at 95\% CL.

\subsection{Flavor-violating $\tau$ decays}

\begin{figure}[t]
	\center{\includegraphics[width=1\textwidth]{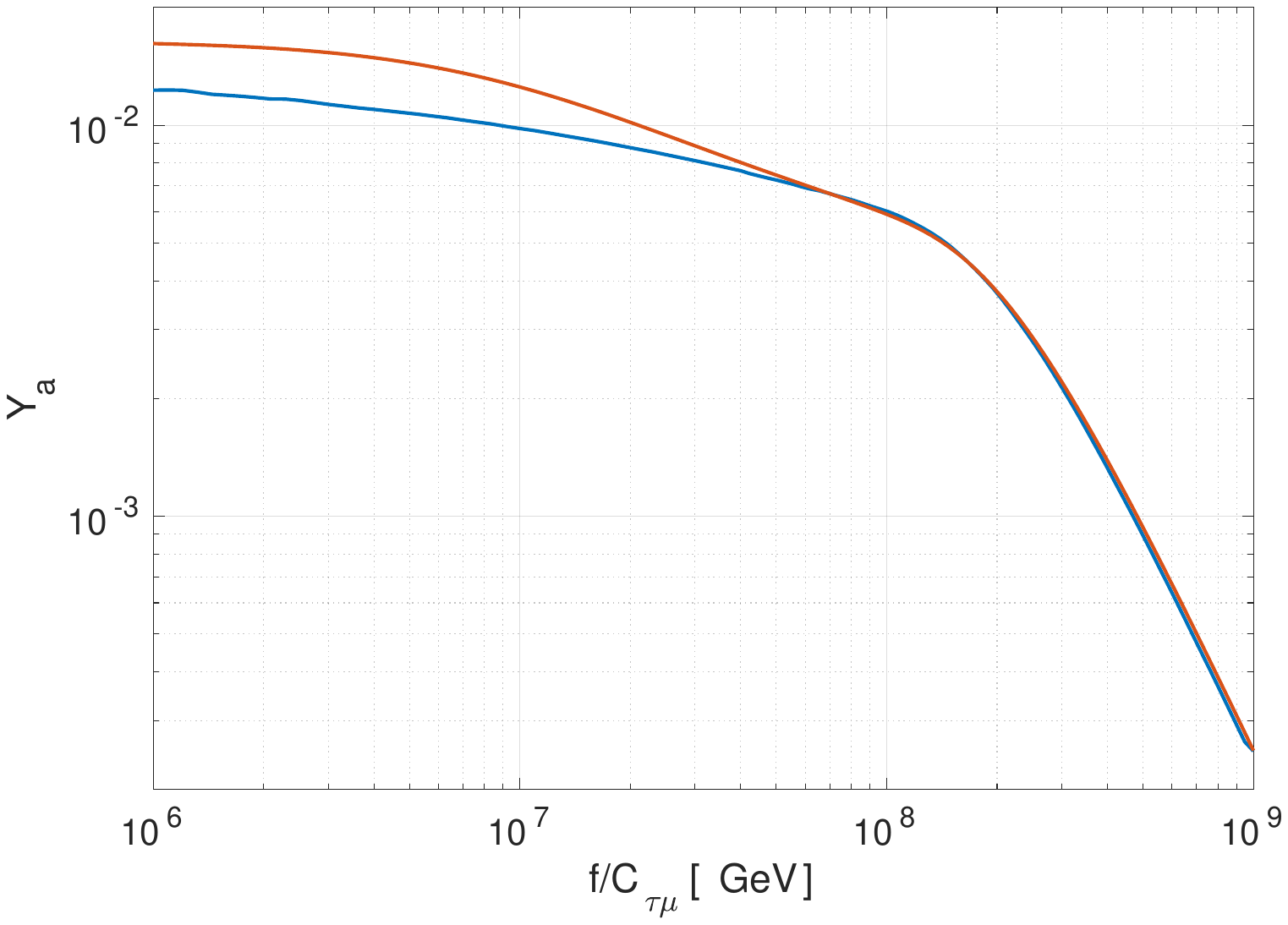}}
	\caption{Comoving relic density of axions $Y_a$ as a function of $f/C_{\mu\tau}$ for non-diagonal coupling of axions to tau. The red curve shows the solution of the standard nBE, the blue curve shows the density from the fBE solution (we use the same color convention in all of the plots below). The initial condition for both approaches is $Y_a(x_{\rm in}) = 0$, where $x_{\rm in} = 0.1$; the evolution was stopped at $x_{\rm f} = 30$ i.e.~well after the final abundance was established.}
	\label{fig:Ya_fa_dec_tau}
\end{figure}

\begin{figure}[t]
	\begin{minipage}[h]{0.49\linewidth}
		\center{\includegraphics[width=1\textwidth]{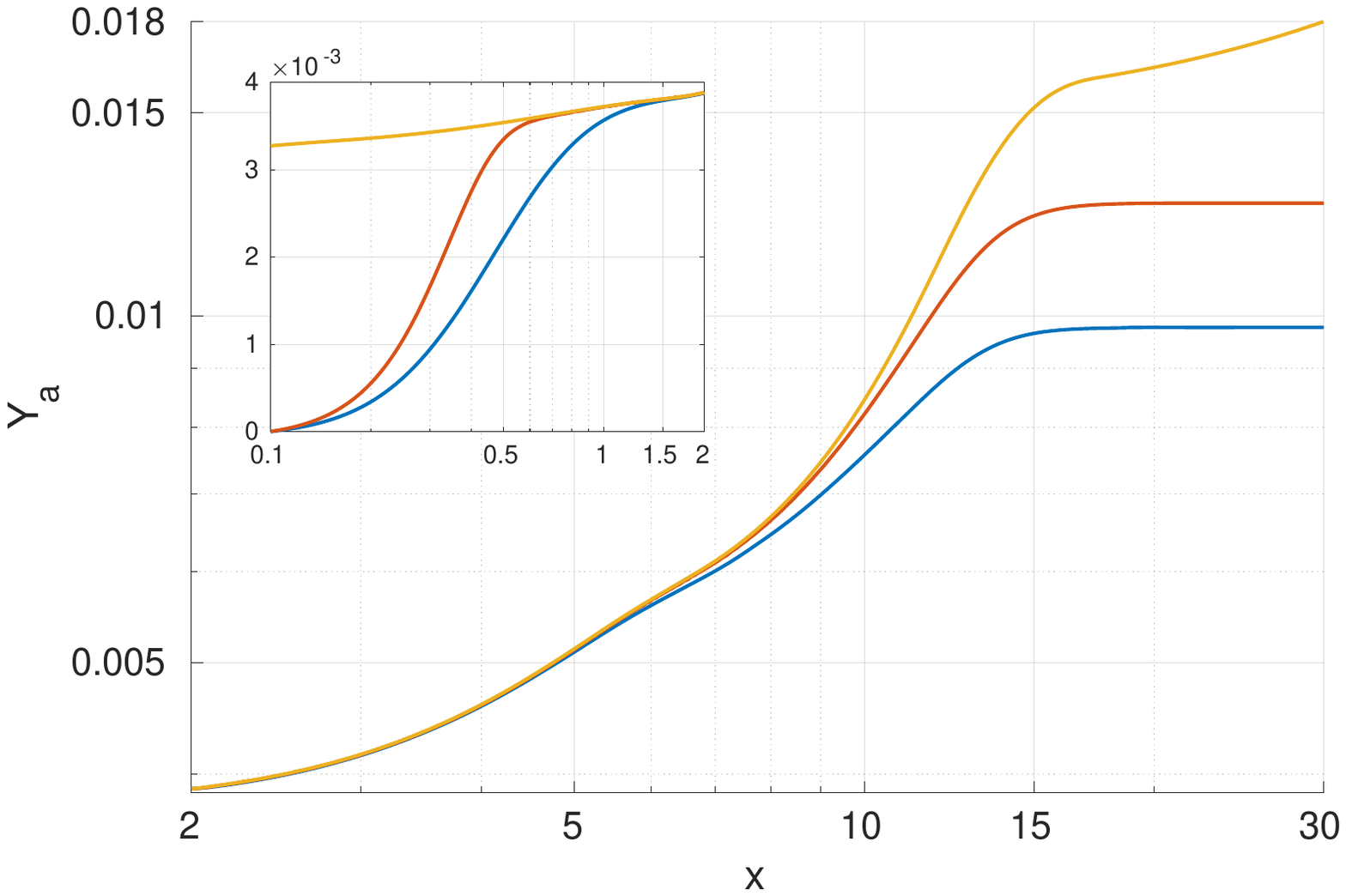}}
	\end{minipage}
	\begin{minipage}[h]{0.5\linewidth}
		\center{\includegraphics[width=1\textwidth]{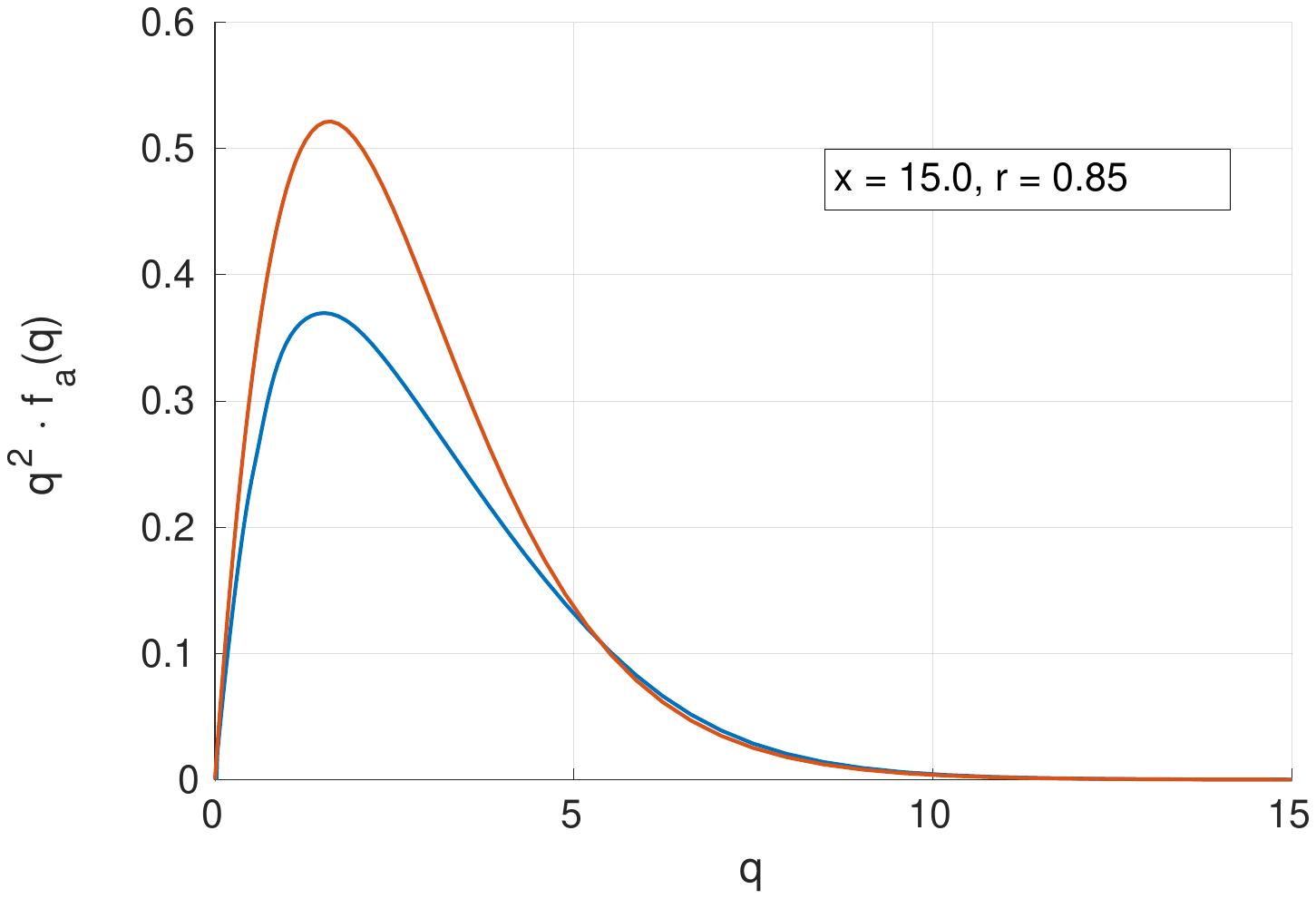}}
	\end{minipage}
	\caption{The comoving density of axions $Y_a$ vs. $x \equiv m_{\tau}/T$ (left) and the distribution function times $q^2$ vs. $q$ (right) for $\tau \rightarrow \mu a$ decays at $f/C_{\tau\mu} = 10^7$ GeV. The blue curve corresponds to the full fBE solution, while the red curve corresponds to the nBE approach. The yellow curve shows the equilibrium density of axions. The right subplot depicts the distribution functions at $x = 15$. The $r$ value denotes the ratio of the energy density computed from the blue distribution to the energy density computed from the red distribution (see text). The smaller plot in the upper left corner of the left plot zooms in on the beginning of the density evolution.} 
	\label{fig:1E+7FO}
\end{figure}

We start with axion production via flavor-violating $\tau$ decays. We present the results for $\tau\to\mu a$ but we checked that the results for $\tau\to e a$ are the same after the substitution $C_{\tau\mu}$ to $C_{\tau e}$.
Fig.~\ref{fig:Ya_fa_dec_tau} shows the comoving relic density of axions for this case predicted by the standard nBE solution (\textit{red}) and the more general solution of the fBE (\textit{blue}) as a function of $f/C_{\tau\mu}$. In this plot and in all of the plots below we use the same colors for the curves obtained with these two approaches. We start the evolution of axion density at $x_{\rm in} = 0.1$ and assume that there are no axions at that moment. We discuss the impact of the choice of initial conditions on our results at the very end of this subsection. The difference between the two approaches is as large as few tens of percent when the axions reach thermal equilibrium and gets close to around $5\%$ when axion production is in the freeze-in regime. 
In cases when axions depart from thermal equilibrium (freeze out) the low-momentum modes decouple from the plasma earlier as the tau decays and inverse decays mostly involve high-momentum modes. 
This is because most of the axion energy in the decay is determined by the mass of the tau lepton, which is getting larger and larger w.r.t. the temperature of the plasma as the latter goes down. The contribution of the low-momentum modes gets suppressed by the rapid change of the entropy degrees of freedom during decoupling corresponding to $x$ between about $7$ and $15$ (compare the distributions in the right subplot of Fig.~\ref{fig:1E+7FO}), hence the resulting abundance of axions for the fBE approach is smaller than for the nBE approach.

\begin{figure}[t]
	\begin{minipage}[h]{0.5\linewidth}
		\center{\includegraphics[width=1\textwidth]{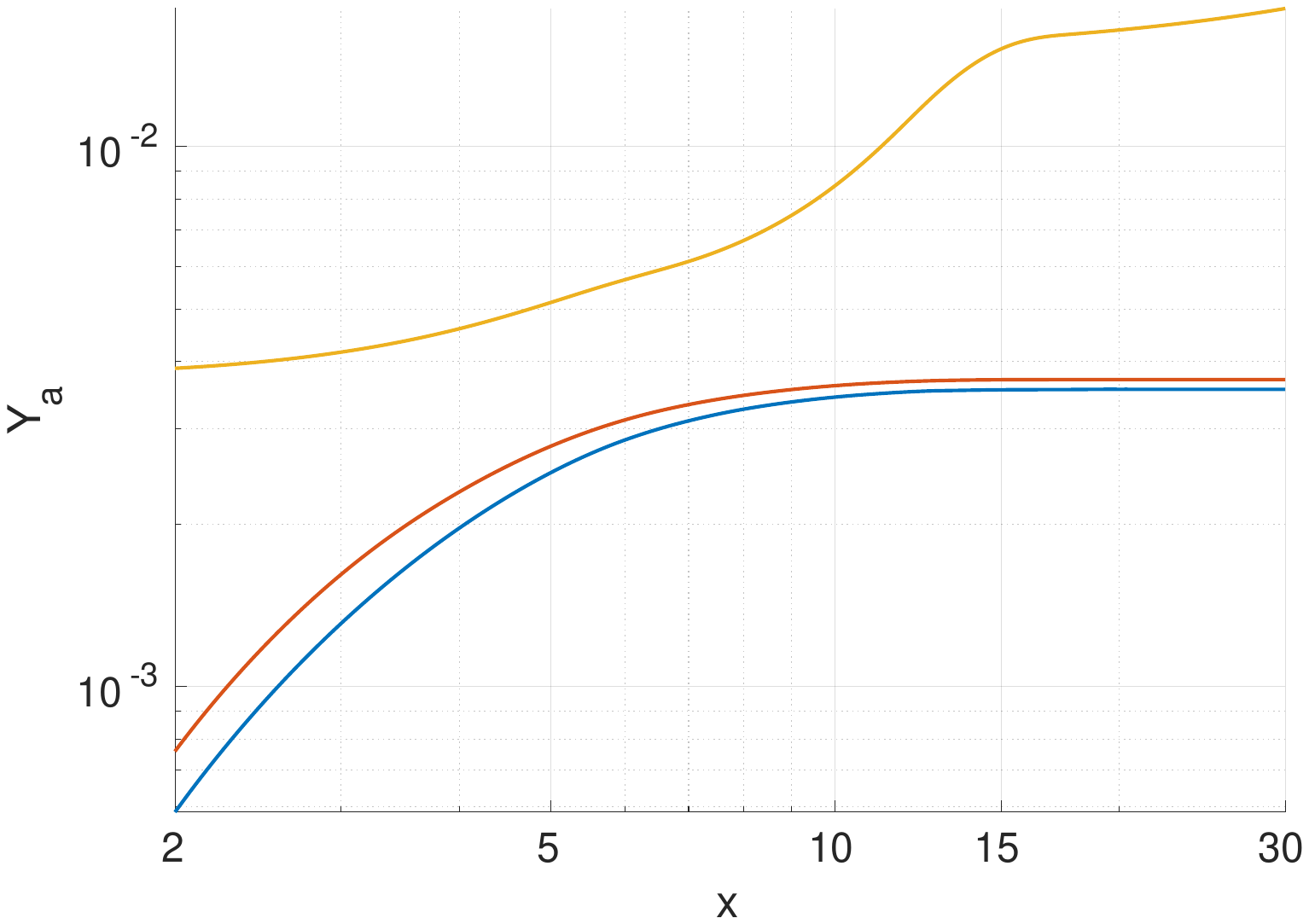}}
	\end{minipage}
	\begin{minipage}[h]{0.5\linewidth}
		\center{\includegraphics[width=1\textwidth]{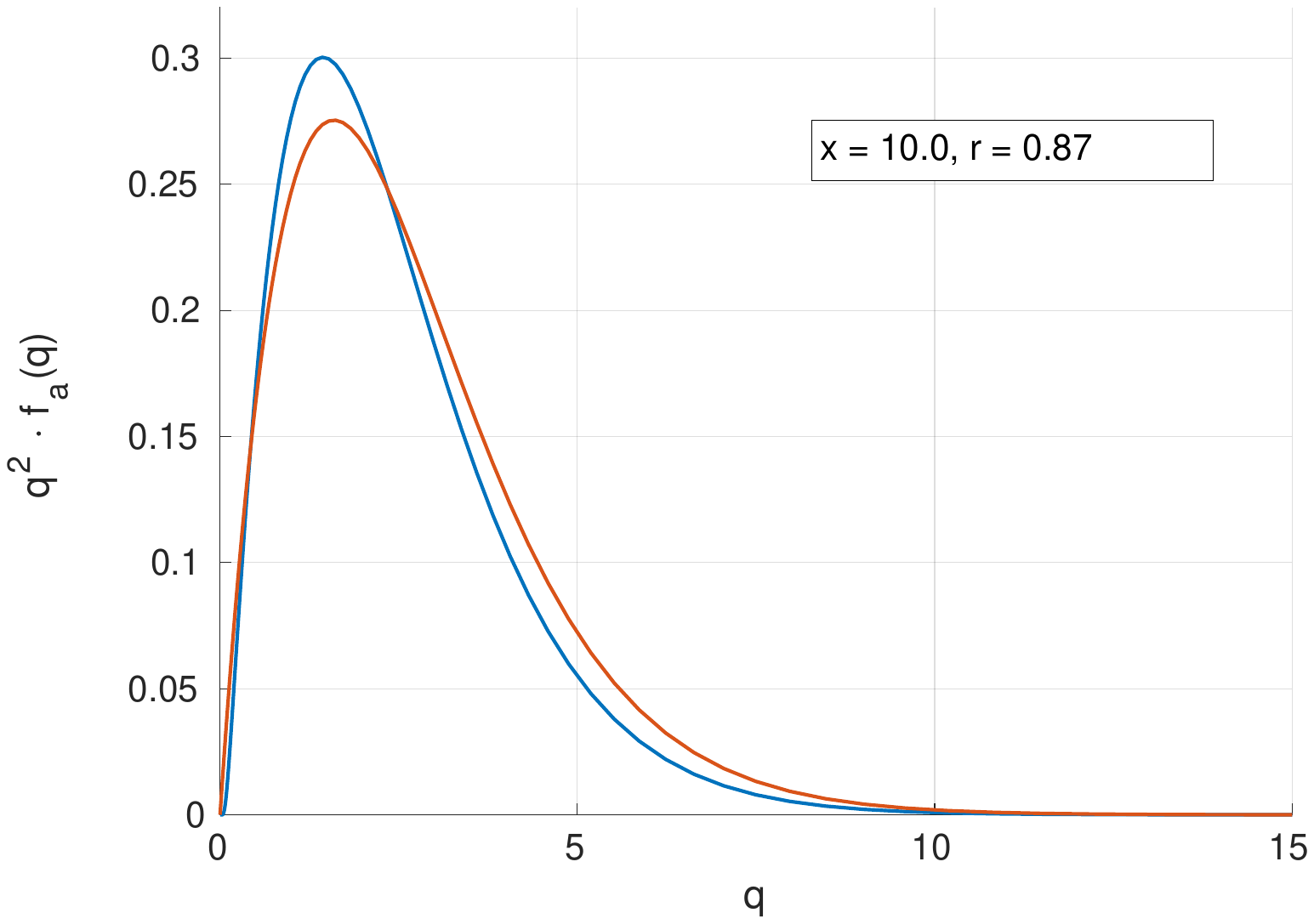}}
	\end{minipage}
     \caption{Same as Fig.~\ref{fig:1E+7FO}, but for $f/C_{\tau\mu} = 2 \cdot 10^8$ GeV. The right subplot depicts the distribution functions at $x = 10$.}
	\label{fig:2E+8FI}
\end{figure}

\begin{figure}[t]
	\center{\includegraphics[width=1\textwidth]{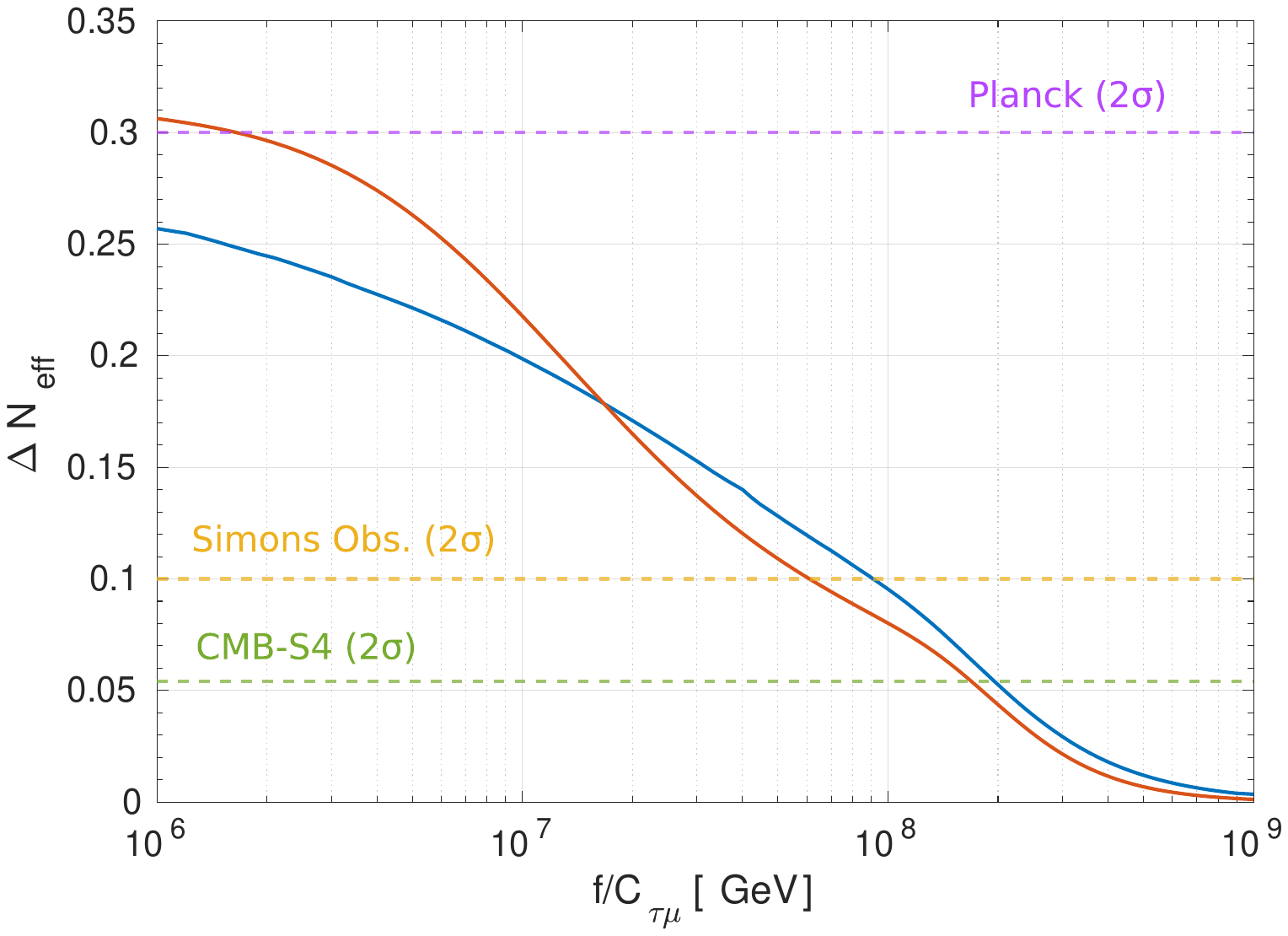}}
	\caption{$\dNeff$ as a function of $f/C_{\mu\tau}$ for non-diagonal coupling of axions to tau leptons and muons. The red curve shows the result for the simplified formula in Eq.~\eqref{eq:dNeff_Simple} with the relic density obtained by the solution of the standard nBE. The blue curve shows the result for the formula in Eq.~\eqref{eq:dNeff_rel_integrated}, where the integral is calculated over the distribution function of axions obtained from the full fBE solution. $\dNeff$ in both cases is calculated at $T \gg m_a$, but when the corresponding value of $h_s = 43/11$. The magenta dashed line indicates the constraint from Planck ($95 \%$ CL) \cite{Planck:2018vyg}. The orange and green dashed lines show the sensitivity (at $2\sigma$ level) of the future experiments Simons Observatory \cite{SimonsObservatory:2018koc} and CMB-S4 \cite{CMB-S4:2016ple}, respectively.
 }
	\label{fig:dNeff_fa_dec_tau}
\end{figure}

In the case of freeze-in the decays dominate over inverse decays and the production rate is almost independent of the axion population. It means that the collision term can be integrated over the axion phase space and the fBE should predict essentially the same number density as the nBE, even though the actual distribution of axions is generally slightly different, see Fig.~\ref{fig:2E+8FI}.

Figs.~\ref{fig:1E+7FO} and~\ref{fig:2E+8FI} demonstrate two examples of the evolution of the comoving density (\textit{left}) and the axion distribution function around decoupling (\textit{right}) for different values of $f$ that correspond to the freeze-out and freeze-in cases. 
Note that in the freeze-out case, in spite of vanishing initial abundance of axions the comoving density reaches the equilibrium density at $x\sim1$ when the axion production rate is maximized (as compared to Hubble expansion rate) as can be seen from a small plot in the left plot of Fig.~\ref{fig:1E+7FO}.
The distribution function for the nBE approach always has a thermal shape (Bose-Einstein) and the normalization that corresponds to the density of axions predicted by the nBE solution for the given temperature. The $r$ factor in the text box in the right subplots indicates the ratio of the energy density computed from the distribution as the fBE solution (\textit{blue}) to the energy density computed from the equilibrium-shaped distribution (\textit{red}) at the given value of $x$. If the axions in both cases evolve further as free particles this ratio should remain constant as long as the axion gas is relativistic and hence translate directly into the ratio of $\dNeff$. However, as the axion decoupling moment cannot be precisely defined this ratio should be rather treated as an estimate of how the difference in the shape of the distribution function affects the energy density. The difference in the shapes of distribution functions in the right subplot of Fig.~\ref{fig:1E+7FO} is explained in the text above where we discuss the difference in the comoving abundances. As for Fig.~\ref{fig:2E+8FI}, the main reason behind the slight difference of the red and blue curves for the comoving abundances in the left subplot is that the fBE properly accounts for the backreaction from the muon plasma (quantum corrections). The deviation of the actual distribution from the thermal shape (right subplot) which is skewed towards smaller momenta is due to a combination of effects stemming from decay kinematics, quantum corrections and the rapid change of entropy degrees of freedom. 

The results for $\dNeff$ predicted by the two approaches are shown in Fig.~\ref{fig:dNeff_fa_dec_tau}. The red curve is calculated from Eq.~\eqref{eq:dNeff_Simple} with $Y_a$ obtained by solving the nBE, while the blue curve is calculated from Eq.~\eqref{eq:dNeff_rel_integrated} with $f_a$ obtained as a solution of the fBE. In the freeze-out regime ($f/C_{\tau\mu} \lesssim10^7$~GeV) the simplified formula for $\dNeff$ serves as a rather good (but not perfect) approximation and predicts a larger value of $\dNeff$ for the nBE solution, because the nBE gives a larger value of the axion relic density. Note, however, that the actual shape of the distribution function is different from the thermal one and features more high-momentum modes for the given number density, cf. the right plot\footnote{The distributions in that plot should not be compared directly as they correspond to different normalizations, though one can deduce from their shapes that were they normalized equally the blue distribution would be skewed more towards the higher momenta.} in Fig.~\ref{fig:1E+7FO}. Thus, if the simplified formula was used for the corresponding value of the comoving number density it would underestimate $\dNeff$. 
For the same reason the fBE solution predicts larger $\dNeff$ for $f/C_{\tau\mu} \gtrsim2\times10^7$~GeV even though the number density from the fBE solution is smaller than that from the nBE solution for $f/C_{\tau\mu}$ up to about $5\times10^7$~GeV, cf.~Figs.~\ref{fig:Ya_fa_dec_tau} and \ref{fig:dNeff_fa_dec_tau}. Still, deep in the freeze-out regime, i.e. for smaller $f/C_{\tau\mu}$, the effect of the suppressed  number density dominates and $\dNeff$ is smaller by up to $0.05$ as compared to the simplified computation \footnote{In the freeze-out regime the axion density during the evolution is large enough to contribute to the effective number of relativistic degrees of freedom. Incorporating this effect into the fBE is rather complicated as one has to solve a coupled system of integro-differential equations, however the estimates such as the one given by Eq.~D.23 in Ref.~\cite{DEramo:2021lgb} yield a correction to the $\dNeff$ that is of order $1/(1 + h_s(T_d))$, where $T_d$ is the temperature of axion decoupling. Hence, in case of tau decays the predicted $\dNeff$ should be larger by $\sim 2 - 6 \%$. Furthermore, the nBE approach should get a larger contribution as axions decouple later in this case.}. This significantly weakens the Planck constraint on $f/C_{\tau\mu}$ from $2\times10^6$~GeV obtained using the simplified approach to well below $10^6$~GeV. This makes the current Planck constraint not competitive to the recent collider constraint from Belle-II~\cite{Belle-II:2022heu}.   

When the axion production processes are getting further from ever reaching the equilibrium, Eq.~\eqref{eq:dNeff_Simple} underestimates the result for $\dNeff$ by a factor of $A^{1/3}$, where $A$ is the normalization factor of the equilibrium distribution of axions (see Eq.~\eqref{eq:axion_density_A} and the text around it) and is significantly lower than unity for frozen-in axions. In this case the shape of the distribution has also an impact on $\dNeff$ but, in contrast to the freeze-out regime, the actual distribution is typically skewed towards smaller momenta of axions (see the right plot in Fig.~\ref{fig:2E+8FI}) which somewhat reduces $\dNeff$ as compared to the thermal shape. Still, the overall enhancement of $\dNeff$ by a factor of $A^{1/3}$ is more important, so for large f the fBE solution predicts larger $\dNeff$ than the standard approach.
Note that if Eq.~\eqref{eq:dNeff_Simple} would correctly account for the relation between the energy density and number density of axions the red curve would always be above the blue curve. Our precise computation of $\dNeff$ improves the sensitivity of future CMB probes. The reach of the Simons Observatory for $f/C_{\tau\mu}$ is improved from about $6\times10^7$ to  $9\times10^7$~GeV. CMB-S4 may improve the bound up to $2\times10^8$~GeV. The prospects for probing $f/C_{\tau\mu}$, as well as $f/C_{\tau e}$, using cosmological measurements are much better than the sensitivity of Belle-II with 50 ab$^{-1}$ of integrated luminosity which is expected to constrain these couplings only up to about $2\times10^7$~GeV~\cite{Calibbi:2020jvd,Badziak:2024szg}.

In our calculations we assumed vanishing initial abundance of axions so let us briefly discuss the impact of different initial conditions. It is well known that at sufficiently high temperatures axion interactions with gluons and/or photons bring axions into equilibrium. When the Universe cools down these interactions cease to be efficient and axions decouple from the SM plasma. As far as axion-photon interactions are concerned axions decouple typically above the electroweak phase transition leading to $\dNeff\approx0.027$, see~e.g.~Ref.~\cite{Green:2021hjh}. Axion-gluon interactions may in principle keep axion in thermal equilibrium down to smaller temperatures, especially for $f\lesssim10^8$~GeV, but below the EW scale the axion-gluon interaction rate suffers from large uncertainties due to non-perturbative effects, see Ref.~\cite{Notari:2022ffe} for a recent discussion on this issue. Since we are interested in conservative bounds on axion couplings to leptons we checked how our predictions for $\dNeff$ change assuming initial population of thermally distributed axions with the abundance corresponding to $\dNeff=0.027$. For small enough $f$, axion-lepton interactions thermalize axions so the final result for $\dNeff$ does not depend on initial conditions. However, in the freeze-in regime, which in case of $\tau$ decays corresponds to $f/C_{\tau\mu}\gtrsim10^8$~GeV, the initial abundance of axions slightly increases the final result for $\dNeff$. For the benchmark point with $f/C_{\tau\mu}=2\times10^8$~GeV, $\dNeff$ increases only by about $0.01$, from about $0.05$ to $0.06$. The impact of the initial abundance of axions on future CMB-S4 sensitivity to $f/C_{\tau\mu}$ is rather small moving the future $2\sigma$ bound from about $2\times10^8$~GeV to about $3\times10^8$~GeV. The impact of different initial conditions on the current Planck bound and the sensitivity of Simons Observatory is completely negligible since the corresponding values of $f/C_{\tau\mu}$ are in the freeze-out regime.

\subsection{Flavor-conserving interactions with leptons}
\label{Sec:results_diagonal}

\begin{figure}[t]
	\begin{minipage}[h]{0.5\linewidth}
		\center{\includegraphics[width=1\textwidth]{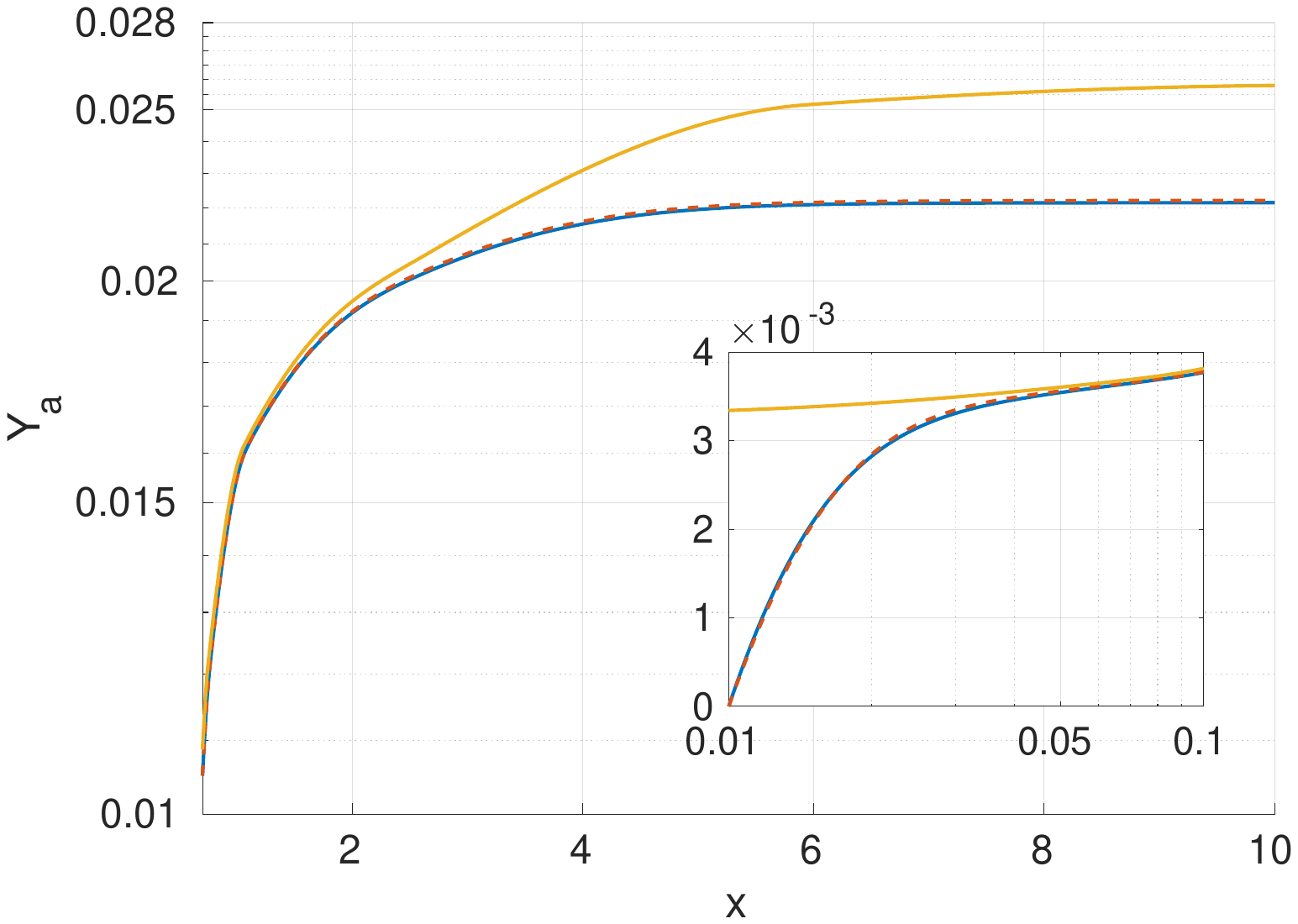}}
	\end{minipage}
	\begin{minipage}[h]{0.5\linewidth}
		\center{\includegraphics[width=1\textwidth]{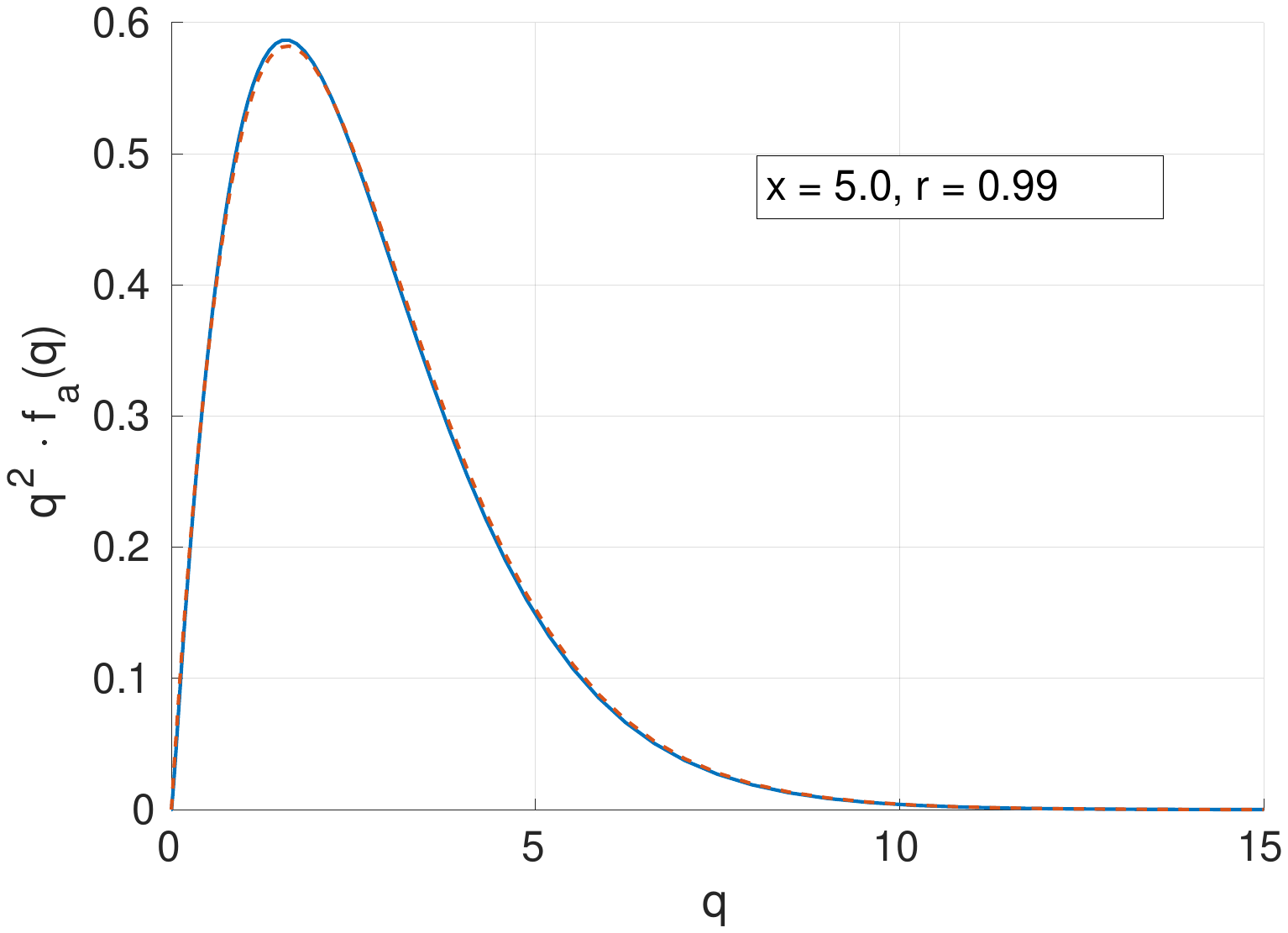}}
	\end{minipage}
	\caption{Same as Fig.~\ref{fig:1E+7FO}, but for processes with muons and $f/C_{\mu} = 10^6$ GeV. The right subplot depicts the distribution functions at $x = 5$. The red curved which corresponds to the standard nBE solution is dashed to make the blue overlapping curve more visible. Note that the equilibrium density curve (yellow) is different than in corresponding plots for tau decays in Figs.~\ref{fig:1E+7FO} and~\ref{fig:2E+8FI} because we use a different definition of $x \equiv m_{\mu}/T$. The initial condition for both approaches is $Y_a(x_{\rm in}) = 0$, where $x_{\rm in} = 0.01$; the evolution was stopped at $x_{\rm f} = 30$ i.e.~well after the final abundance was established. The smaller plot in the lower right corner of the left plot zooms in on the beginning of the density evolution.
 }
	\label{fig:1E+6_muons}
\end{figure}

\begin{figure}[t]
	\begin{minipage}[h]{0.5\linewidth}
		\center{\includegraphics[width=1\textwidth]{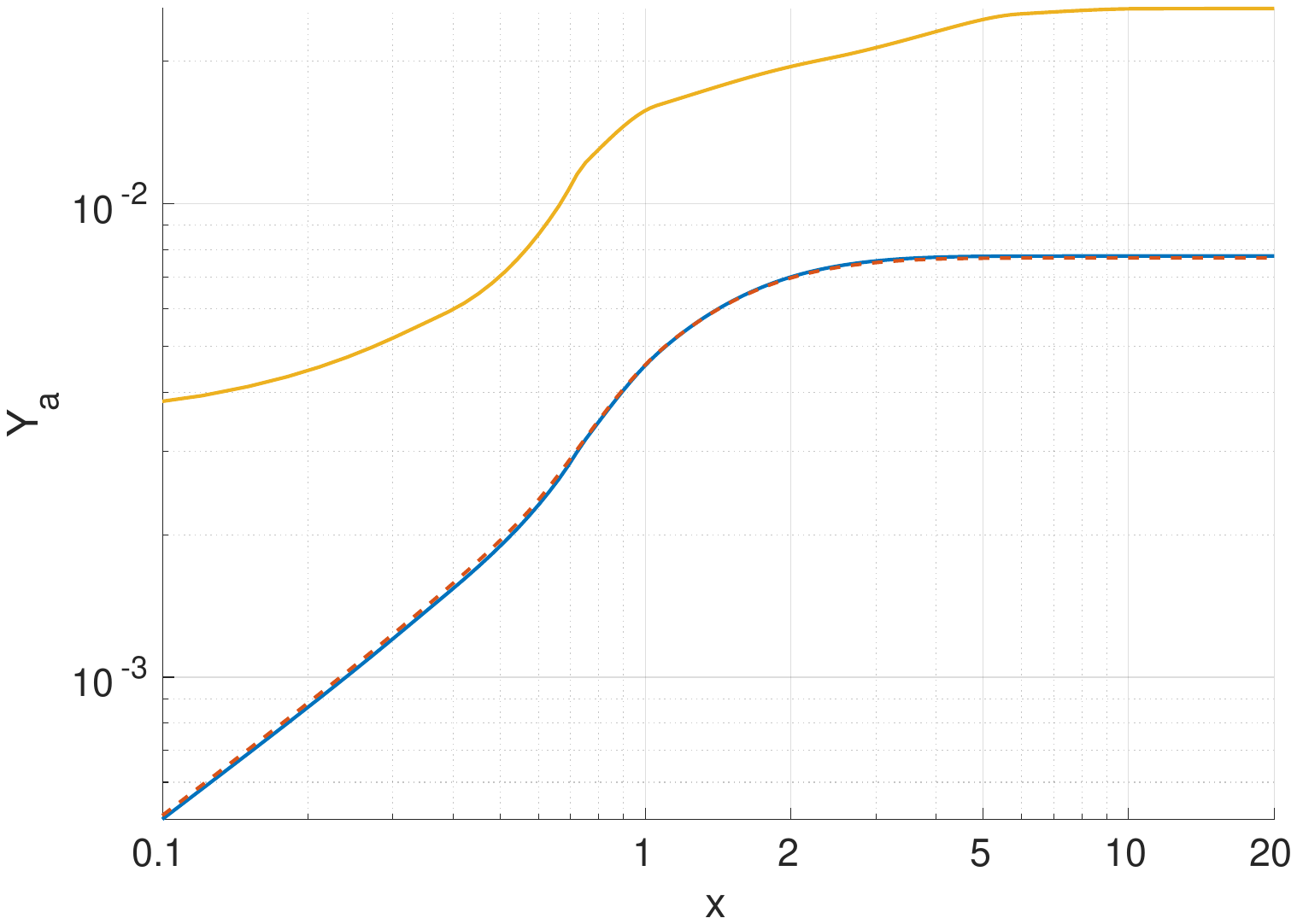}}
	\end{minipage}
	\begin{minipage}[h]{0.5\linewidth}
		\center{\includegraphics[width=1\textwidth]{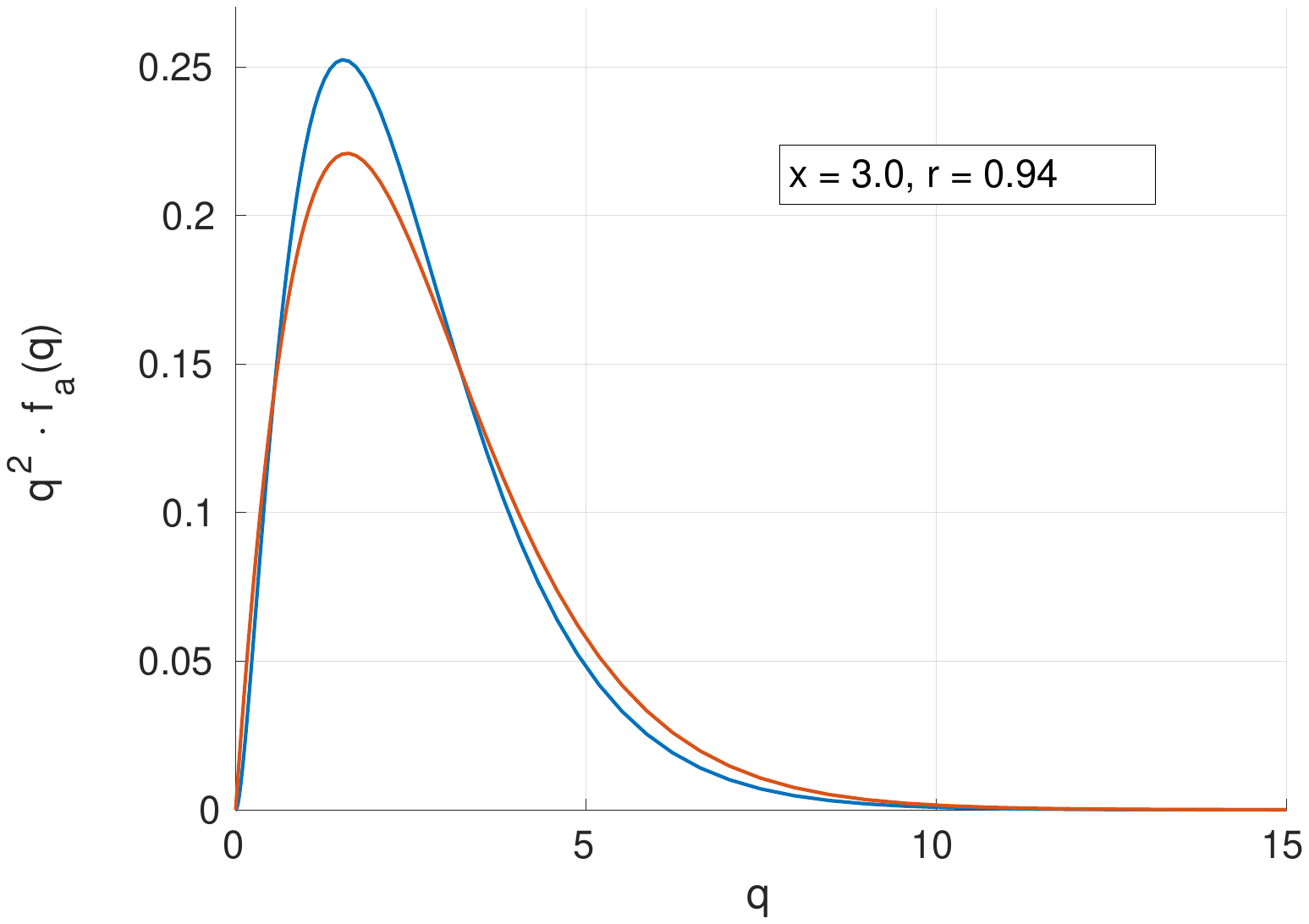}}
	\end{minipage}
	\caption{Same as Fig.~\ref{fig:1E+6_muons}, but for $f/C_{\mu} = 10^7$ GeV. The right subplot depicts the distribution functions at $x = 3$. The red curve in the left subplot is dashed to make the overlapping blue curve more visible.}
	\label{fig:1E+7_muons}
\end{figure}

Let us now focus on the axion produced via flavor-conserving annihilations and scatterings with muons and tau leptons. For this type of processes we start the evolution of the density (or distribution function) at $x_{\rm in} = 0.01$. We found that in the whole range of $f$ values both approaches (nBE and fBE) predict essentially the same axion density evolution. This is demonstrated in Figs.~\ref{fig:1E+6_muons} and ~\ref{fig:1E+7_muons} for two representative values of $f/C_{\mu}$ equal $10^6$ and $10^7$~GeV, corresponding to the freeze-out and freeze-in axion production via interactions with muons, respectively. We found the same behavior for the case of axion-tau scattering. The difference between the axion relic densities given by these approaches is the smallest when axions reach thermal equilibrium and increases with $f$ (as the axion population is further and further away from reaching the equilibrium), but generally remains below $5\%$. This is expected, because the energy of axions produced by flavor-diagonal interactions is mostly defined by the temperature of plasma, especially for the Primakoff scattering, and the deviation from the thermal shape of the distribution should be smaller than in the case of decays. 

\begin{figure}[t]
    \begin{minipage}[h]{0.5\linewidth}
	\center{\includegraphics[width=1\textwidth]{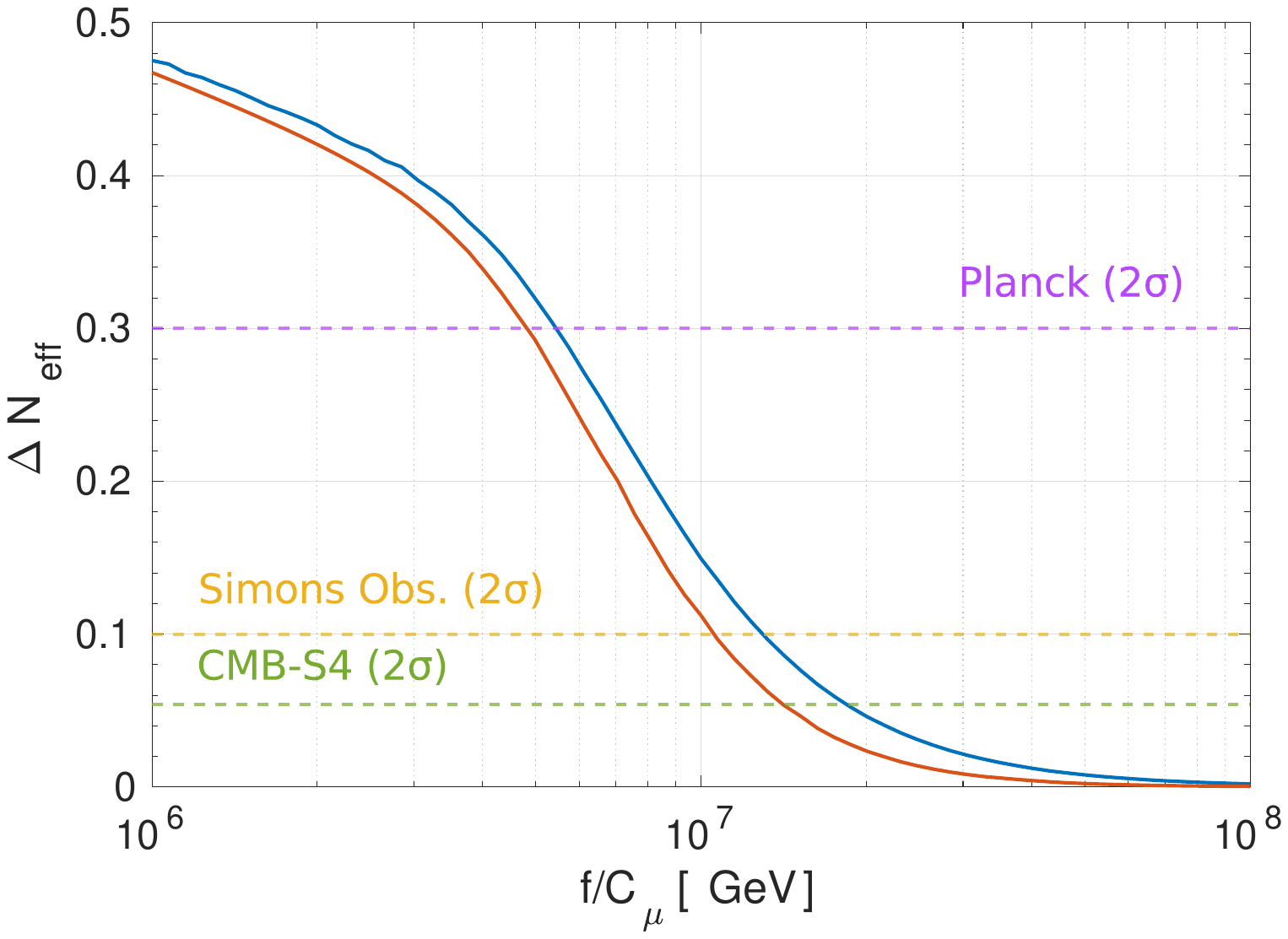}}
    \end{minipage}
    \begin{minipage}[h]{0.5\linewidth}
	\center{\includegraphics[width=1\textwidth]{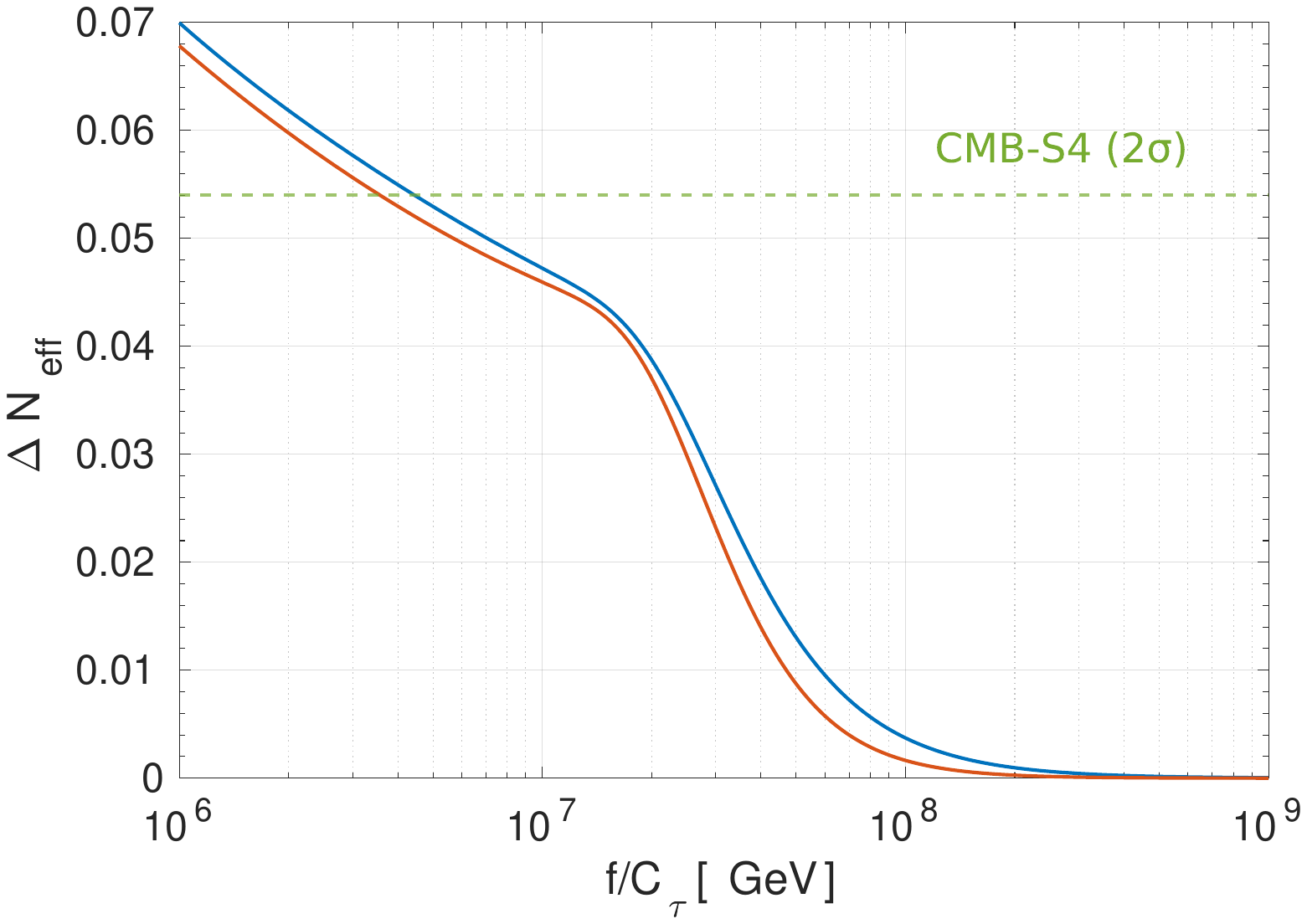}}
    \end{minipage}
	\caption{Same as Fig.~\ref{fig:dNeff_fa_dec_tau}, but for diagonal coupling of axion to muons (\textit{left}) and tau leptons (\textit{right}).  
    }
	\label{fig:dNeff_fa_muons_tau}
\end{figure}

Since the axion number density predicted by the two approaches is essentially the same the difference in $\dNeff$ shown in Fig.~\ref{fig:dNeff_fa_muons_tau} (\textit{left}) has its origin in the simplified formula~\eqref{eq:dNeff_Simple} for $\dNeff$ used in the nBE approach. In the freeze-out regime, corresponding to values of $f/C_{\mu}\sim\mathcal{O}(10^6)$~GeV, the difference is quite small and in any case $\dNeff$ exceeds the upper bound from Planck. In the freeze-in regime, the situation is qualitatively similar to freeze-in axion production via tau decays so $\dNeff$ is enhanced due to the unaccounted factor $A^{1/3}$ in the simplified formula with a subdominant correction due to the different shape of the actual distribution of axions. The enhancement is bigger than in the case of tau decays. For $f/C_{\mu}$ around $10^7$~GeV  $\dNeff$ is increased by about $0.05$ which is comparable to $2\sigma$ uncertainties expected from CMB-S4.

We perform the same analysis of flavor-diagonal interactions for axion production via annihilations and scatterings of tau leptons. The results for $\dNeff$ are presented in Fig.~\ref{fig:dNeff_fa_muons_tau} (\textit{right}), while the predicted $Y_a$ for the two approaches are essentially the same in similarity with the muon case. The abundance of axions produced at the tau-lepton temperature scale is suppressed w.r.t the typical axion abundance produced at the muon temperature scale due to the difference in the entropy degrees of freedom and hence the observational perspectives or constraints on that single production channel are weaker. Still, CMB-S4 may test values of $f/C_{\tau}$ up to about $5\times10^6$~GeV which would be the world-leading limit on this coupling.

Similarly as in the case of $\tau$ decays, the initial conditions for axion abundance do not affect the final result for $\dNeff$ as long as axion-lepton scatterings thermalize axions. However, in the case of scattering the freeze-in regime corresponds to smaller values of $f/C_{\mu}\gtrsim\mathcal{O}(10^7)$~GeV. Assuming that axions were thermalized at high temperatures and decoupled above the electroweak scale, which corresponds to initial axion abundance giving $\dNeff=0.027$, for the benchmark value of $f/C_{\mu}=10^7$~GeV the final result for $\dNeff$ is increased by less than 0.02 as compared to the result assuming vanishing initial thermal abundance. Still, the impact on the sensitivity of CMB experiments is marginal e.g. the future $2\sigma$ bound from CMB-S4 is shifted from slightly below to slightly above $2\times10^7$~GeV. In the case of $\tau$ scatterings the impact of initial conditions on future CMB-S4 sensitivity is completely negligible because it will only probe values of $f/C_{\tau}$ corresponding to the freeze-out regime.

\section{Conclusions and Outlook}
\label{Sec:Conclusion}

A commonly used assumption in the calculation of the energy density of thermally produced axions, parameterized by their contribution to $\dNeff$, is that axions have equilibrium momentum distribution. 
In this work, we have investigated thermal production of QCD axions in the early Universe and computed the axion contribution to $\dNeff$ relaxing the above assumption.
To this end, we have solved Boltzmann equations for the distribution functions in the full phase space including also the corrections from quantum statistics of the particles in the plasma. This approach leads to precise predictions for the axion contribution to $\dNeff$ for any value of the axion decay constant.

We have focused on the axion production via flavor-conserving axion interactions with muons or tau leptons and flavor-violating axion interactions with tau lepton and electron or muon. Laboratory and astrophysical constraints for the above interactions are relatively weak, so the cosmological probes may be the most sensitive to such interactions, which is crucial to have a precise prediction for $\dNeff$ for these production channels. We found the biggest correction to the prediction for $\dNeff$ for the axions produced via axion-muon scattering and flavor-violating $\tau\to l a$, where $l=e\,\rm{or}\,\mu$, which can be up to about $0.05$. On the other hand, precise prediction for $\dNeff$ from axion-tau scatterings is different from the simplified approach result by less than $0.01$. 

Our results have important implications for current and future constraints on the axion couplings to muons as well as flavor-violating axion couplings with tau. In the case of flavor-violating tau decays the current limits on $f/C_{\tau e}$ and $f/C_{\tau \mu}$ from the Planck constraints on $\dNeff$ is significantly relaxed to the level $\mathcal{O}(10^5)$~GeV because our more precise prediction for $\dNeff$ is smaller by $0.05$ than that obtained with the simplified approach in the region in which $\tau$ decays bring axion to thermal equilibrium. As a result, the limits from Planck are much weaker than the bounds recently obtained by Belle-II from direct searches for $\tau\to e a$ and $\tau\to \mu a$ decays. On the other hand, for larger $f$, corresponding to freeze-in production of axions, our results give larger $\dNeff$ which strengthens the expected future limits on $f/C_{\tau e}$ and $f/C_{\tau \mu}$ from Simons Observatory and CMB-S4 to about $10^8$ and $2\times10^8$~GeV, respectively. The expected limits from cosmology are expected to be stronger by an order of magnitude than the future accelerator limits from Belle-II.

In the case of muon scattering, we found $\dNeff$ to be enhanced w.r.t. the standard approach for any value of $f$. The biggest difference occurs for $f/C_{\mu}$ around $10^7$~GeV, where our result is bigger by about $0.05$ which is comparable to future CMB-S4 sensitivity at $2\sigma$ level. Therefore, our results should be used to reliably compare predictions with the results of future CMB experiments.

The underlying physics behind the difference in the predictions for $\dNeff$ between our precise computation and the simplified one depends on whether the axion is in thermal equilibrium with the SM plasma or the production occurs via freeze-in. In the case of axions thermalized by flavor-violating tau decays, the low-momentum modes decouple from the SM plasma earlier which results in smaller number density of relic axions due to the rapidly decreasing entropy degrees of freedom. On the other hand, in the freeze-in regime, corresponding to larger axion decay constants, axion momentum distribution has a non-thermal shape and the relation between energy density and number density of axions used in the simplified approach is no longer a good approximation leading to an underestimation of $\dNeff$ as compared to our precise result. The underestimation of the standard approach is the most significant for axion-muon scattering.

Our method can be straightforwardly applied to the axion production via interactions with quarks. However, prediction for the axion production rate from scatterings with quarks suffers from large uncertainties from non-perturbative effects which are larger than the uncertainties from the use of the simplified approach to the solution of Boltzmann equations, see Ref.~\cite{Notari:2022ffe} for a recent discussion.  

The present work opens new avenues for further research. In our analysis, we assumed that only one type of axion coupling to SM particles is present at a time. In many axion models several axion production channels can be relevant and it would be interesting to study the interplay between various channels and the impact on $\dNeff$ by solving Boltzmann equations in the momentum space. In particular, it would be interesting to refine predictions for $\dNeff$ in astrophobic axion models~\cite{Badziak:2024szg} or models in which axion production occurs via both axion-pion scattering and axion interactions with SM leptons such as DFSZ models~\cite{Ferreira:2020bpb}.

Another interesting direction is to investigate cosmological constraints on axion couplings taking into account the effect of the non-vanishing axion mass. Such effects are expected to be relevant for values of the axion decay constant around $10^7$~GeV and below, for which the axion mass is comparable to the recombination temperature so the axions cannot be treated as ultra-relativistic at the CMB scale, see~e.g.~\cite{Caloni:2022uya}. Distortions of the axion momentum distribution found in the present paper may also affect the cosmological constraints. We plan to investigate the above matters in the future.

\bigskip

\textbf{Note added:} a few days after we published the pre-print of this work on \texttt{arXiv}, Ref.~\cite{DEramo:2024jhn} appeared on \texttt{arXiv} which studied thermal axion phase-space evolution for flavor-diagonal axion interactions with leptons and quarks.

\section*{Acknowledgments} 

The authors would like to thank Michał Łukawski and Seokhoon Yun for useful discussions. This work was partially supported by the National Science Centre, Poland, under research grant no. 2020/38/E/ST2/00243. 

\appendix

\section{Collision term and rate for 2-body decay into an axion}
\label{apx:decay}

Let us consider the 2-body decay into an axion on the example of the $\tau \rightarrow \mu a$ process that we consider in this work. The collision term for this decay is given by 

\begin{multline}
C[f_a] = \frac{1}{2g_aE_a} \int \frac{d^3p_\tau}{(2\pi)^32E_\tau} \int \frac{d^3p_\mu}{(2\pi)^32E_\mu} \; (2\pi)^4 \delta^{(4)}(\mathcal{P}_\tau - \mathcal{P}_\mu - \mathcal{P}_a) \, \MM_{\tau \rightarrow \mu a} \times \\ \times 2 \Big[ f_\tau (1 + f_a)(1 - f_\mu) - f_\mu f_a(1 - f_\tau) \, \Big] \, ,
 \label{eq:CollDecGen}
\end{multline} 
where $g_a$ is the number of degrees of freedom\footnote{Note that the number of degrees of freedom of particles in the reaction do not enter the collision term on a par with distribution functions (and are not included into them). They are only used to recover the physical number density of these states. Otherwise, collision term will not be equal to zero in full equilibrium.} of particle $a$.
The first term in this formula describes the impact of the decay of $\tau$, while the second term corresponds to the inverse decay of $\tau$. The factor of $2$ in front of the square brackets takes into account the contribution to the axion production or destruction from leptons and antileptons.

To simplify this collision term in full generality let us take a look at the part of the expression above in square brackets that consists of distribution functions

\be
f_{\tau}(1+f_a)(1-f_{\mu}) - f_a f_{\mu}(1-f_{\tau}) = f_{\tau}(1+f_a) - f_{\tau}f_{\mu} - f_a f_{\mu} \, .
\label{eq:decay_dist_functions}
\ee
If axions would be in equilibrium (given that muons and tau are always in thermal equilibrium with the plasma) the expression above should be zero, because it is the only condition for the collision term to be 0 in equilibrium. Hence, we can derive the expression for the second term 
\be
f_{\tau}f_{\mu} = f_{\tau}(1+f^{\rm eq}_a) - f^{\rm eq}_a f_{\mu} \, , 
\label{eq:equilibrium_taumu_term}
\ee
so Eq.~\eqref{eq:decay_dist_functions} takes the following simple form
\begin{equation} 
(f^{\rm eq}_a - f_a)(f_{\mu} - f_{\tau})  \, .
\end{equation} 
With this the decay collision term can be rewritten as (see also Eq.~\eqref{eq:collterm_general_axion})

\be
C[f_a] = \frac{1}{2g_aE_a} \left( 1 - \frac{f_a}{f^{\rm eq}_a} \right) \gamma_{\rm dec} \, , .
\label{eq:collterm_general_decay}
\ee
with all the integrals contained in the $\gamma$ function 

\begin{equation}
\gamma_{\rm dec} = 2 f^{\rm eq}_a \int \frac{d^3p_\tau}{(2\pi)^32E_\tau} \int \frac{d^3p_\mu}{(2\pi)^32E_\mu} \, (2\pi)^4 \delta^{(4)}(\mathcal{P}_\tau - \mathcal{P}_\mu - \mathcal{P}_a) \, \MM_{\tau \rightarrow \mu a} (f_{\mu} - f_{\tau})
\label{eq:gamma_dec_full}
\end{equation}
The amplitude squared for this decay (summed over final \textbf{and} initial polarizations)

\begin{equation}
\MM_{\tau \rightarrow \mu a} = \lp \frac{C_{\tau\mu}}{f} \rp^2 \frac{m^4_{\tau}}{2} \lp 1 - r^2 \rp^2 \, ,
\end{equation}
where $r = m_{\mu}/m_{\tau}$, does not depend on any kinematical variables and can be factorized outside of the integral. The integral can be solved analytically term by term without any further assumptions leading to Eqs.~\eqref{eq:gamma_dec_main}, \eqref{eq:xi_dec_1} and \eqref{eq:xi_dec_2}, where we also used the analytical relation between the squared amplitude and the decay width $\Gamma$ (see below). 

To retrieve the standard approach (Gelmini-Gondolo) of solving the number density Boltzmann equation (nBE) directly one should integrate the Eq.~\eqref{eq:FullBoltzmann} over the axion phase-space, as described in Sec.~\ref{Sec:fBE}. The left-hand side of the fBE can be integrated straightforwardly, so the first term simply becomes $\tilde{H}x\partial_xn_a$ and the second term integrated by parts gives $0$ for any realistic distribution function and independently of $\tilde{g}$. Integrating the collision term in the same way is much more complicated and an analytic expression can be derived under certain approximations. 

First of all, the collision term contains an unknown function $f_a$, hence we need to introduce some ansatz for it to perform the integration. The common choice here is $f_a = (n_a/n^{\rm eq}_a) f^{\rm eq}_a$, which means that the distribution of axions has the \textit{equilibrium} shape, although the actual number density $n$ can deviate from the equilibrium value $n^{\rm eq}$. This regime can be maintained by the elastic scattering of axions on the SM particles in the plasma (kinetic equilibrium), however as axion elastic scattering are suppressed in magnitude at best by a factor of $(1/f)^2$ w.r.t. axion production processes the precision of this approximation is questionable. 

The second approximation that greatly simplifies the integration of the collision term is neglecting the quantum corrections, i.e. assuming that $(1 \pm f_i) \approx 1$, which is reasonable for non-relativistic or extremely diluted gases. For particles in thermal equilibrium such as muons or tau leptons this is equivalent to using the MB distribution. Combining both approximations and using the same equilibrium identity as in the derivation of Eq.~\eqref{eq:equilibrium_taumu_term} transforms Eq.~\eqref{eq:decay_dist_functions} into $f_{\tau} (1 - n_a/n^{\rm eq}_a)$ and hence allows us to formulate the integrated collision term directly\footnote{We managed to write the collision term via $\Gamma_{\tau \rightarrow \mu a}$ (in Eq.~\ref{eq:gamma_dec_main} ) only because $\MM_{\tau \rightarrow \mu a}$ does not have any angular dependencies. } via the decay width $\Gamma$
\be
\int \frac{d^3p_a}{(2\pi)^32E_a} \int \frac{d^3p_\mu}{(2\pi)^32E_\mu} (2\pi)^4 \delta^{(4)}(\mathcal{P}_\tau - \mathcal{P}_\mu - \mathcal{P}_a) \MM_{\tau \rightarrow \mu a} = 2 g_\tau m_\tau \Gamma_{\tau \rightarrow \mu a} \, .
\label{eq:decay_width}
\ee
The factor of $g_\tau = 2$ appears because the standard definition of the decay width involves $\bar{\MM}$ \textit{averaged} over initial polarizations, while the collision term should account for all the possible reactions that change the number of axions, including those with different polarizations. 

The rhs of the nBE can be now written as 

\be
\bGamma^+_a - \bGamma^-_a = 2 \cdot 2 \, g_\tau m_\tau \Gamma_{\tau \rightarrow \mu a} \left( 1 - \frac{n_a}{n^{\rm eq}_a}\right) \int \frac{d^3p_\tau}{(2\pi)^32E_\tau} f_\tau \, ,
\ee
and due to a factor of $2E_\tau$ in the integrand cannot be directly expressed in terms of $n_{\tau}$. Using MB distribution for tau leptons one gets

\be
\int \frac{d^3p_\tau}{(2\pi)^32E_\tau} \, f_{\tau} \, = \frac{m_\tau T}{4\pi^2} K_1 \left(\frac{m_\tau}{T}\right) \, ,
\ee
while the number density is 

\be
g_\tau \int \frac{d^3p_\tau}{(2\pi)^3} \, f_\tau \, = g_\tau \frac{m^2_\tau T}{2\pi^2} K_2 \left(\frac{m_\tau}{T}\right) \, ,
\label{eq:density_MB}
\ee
where $K_1$ and $K_2$ are the modified Bessel function of the first and second order respectively. Connecting the first relation to the second one via the temperature one finally arrives at the rhs given by Eq.~\eqref{eq:rhs_nbe_decay}.

\section{Collision terms and rates for $2 \rightarrow 2$ processes with an axion}
\label{apx:2to2}

The general collision term for the process $i + j \rightarrow a + k $ is given by 

\begin{multline}
C[f_a] = \frac{1}{2g_aE_a} \int d\Pi_k d\Pi_i d\Pi_j \, (2\pi)^4 \delta^{(4)}(\mathcal{P}_i + \mathcal{P}_j - \mathcal{P}_a - \mathcal{P}_k) \MM_{ij \rightarrow ak} \times  \\
\times \Big[ f_i f_j (1 + f_a)(1 \pm f_k) - f_a f_k (1 \pm f_j)(1 \pm f_i) \, \Big] \, ,
\label{eq:CollAnnGen}
\end{multline} 
where $d\Pi_i = d^3p_i/((2\pi)^32E_i)$ and the integral is taken over the phase space of all the particles except $a$ and the signs of the quantum corrections depend on the spin statistics of the corresponding particles and in general should not have the same value across this expression in all the cases. As we only consider interactions of axions with SM particles in this work $i,j$ and $k$ particles are considered to be in thermal equilibrium. Let us take a look at the distribution function part of the integrand. In full thermal equilibrium the second term is equal to the first one, hence we can express the second term as

\be
f_a f^{\rm eq}_k (1 \pm f^{\rm eq}_i)(1 \pm f^{\rm eq}_j) = \frac{f_a}{f^{\rm eq}_a} f^{\rm eq}_i f^{\rm eq}_j (1 + f^{\rm eq}_a)(1 \pm f^{\rm eq}_k) \, .
\label{eq:EquilIdentityAnn}
\ee
Plugging this back into the collision term we get the same structure as given by Eqs.~\eqref{eq:collterm_general_axion} and \eqref{eq:collterm_general_decay} with 

\be
\gamma_{ij \rightarrow ak} = \int d\Pi_k d\Pi_i d\Pi_j \, (2\pi)^4 \delta^{(4)}(\mathcal{P}_i + \mathcal{P}_j - \mathcal{P}_a - \mathcal{P}_k) \MM_{ij \rightarrow ak} f^{\rm eq}_i f^{\rm eq}_j (1 \pm f^{\rm eq}_k) \, .
\label{eq:SmallGammaAnn}
\ee

Before elaborating this integral any further let us go through some useful quantities and definitions. The most general formula for the \textbf{rate} of the process $i +j \rightarrow a + k$ is

\be
\bGamma_{ij \rightarrow ka} = \int d\Pi_a d\Pi_k d\Pi_i d\Pi_j \, (2\pi)^4 \delta^{(4)}(\mathcal{P}_i + \mathcal{P}_j - \mathcal{P}_a - \mathcal{P}_k) \MM_{ij \rightarrow ak} f_i f_j (1 \pm f_k) (1 + f_a)\, .
\label{eq:AnnRateGen}
\ee
This rate defines the frequency (per unit volume) of $i$ and $j$ transformations into $k$ and $a$ which takes into account that \textit{not any final state is possible} due to the blocking/stimulated emission effects (quantum corrections). In general, this rate is not equal to the rate of the inverse process $\bGamma_{ka \rightarrow ij}$ (only in chemical equilibrium).

The total \textbf{cross section} of this process is defined as follows

\be
\sigma_{ij \rightarrow ka} = \frac{1}{4g_i g_j E_iE_j v_{\rm Mol}} \int d\Pi_k d\Pi_a \, (2\pi)^4 \delta^{(4)}(\mathcal{P}_i + \mathcal{P}_j - \mathcal{P}_a - \mathcal{P}_k) \MM_{ij \rightarrow ak} \, ,
\ee
where $v_{\rm Mol}$ is the Moller velocity

\be
v_{\rm Mol} = \frac{\sqrt{(\mathcal{P}_i \cdot \mathcal{P}_j)^2 - m^2_i m^2_j}}{E_iE_j} = \frac{p_{ij} \sqrt{s}}{E_iE_j} \, ,
\label{eq:MollerVelocity}
\ee
and $p_{ij}$ is the momentum of incoming particle in the CM frame. Note that the numbers of degrees of freedom $g_i$ and $g_j$ are usually incorporated in the $\MM$ \textit{averaged} over initial polarizations, while as discussed below Eq.~\eqref{eq:decay_width} for our purposes we keep track of the $\MM$ summed over initial and final states.  If we \textit{assume} that the quantum corrections $(1+f_i)$ can be neglected, the rate in Eq.~\eqref{eq:AnnRateGen} can be expressed through the cross section as 

\be
\bGamma_{ij \rightarrow ka} \approx g_i g_j \int \frac{d^3p_i}{(2\pi)^3} \frac{d^3p_j}{(2\pi)^3} \sigma_{ij \rightarrow ka} v_{\rm Mol} \, f_i f_j \, ,
\label{eq:AnnRateIntegral} 
\ee 
The expression above resembles the $\sigma_{ij \rightarrow ka} v_{\rm Mol}$ averaged over the full phase space of the initial states with the corresponding weights

\be
\langle \sigma_{ij \rightarrow ka} v_{\rm Mol}\rangle = \frac{\displaystyle \; g_i g_j \int \frac{d^3p_i}{(2\pi)^3} \frac{d^3p_j}{(2\pi)^3} \sigma_{ij \rightarrow ka} v_{\rm Mol} \, f_i f_j \; }{\displaystyle g_i g_j \int \frac{d^3p_i}{(2\pi)^3} \frac{d^3p_j}{(2\pi)^3} \, f_i f_j} \, .
\label{eq:sigmaV_general}
\ee
Since the integrals in the denominator simply give the densities of $i$ and $j$ particles we arrive at the standard expression for the rates used in the nBE in Eq.~\eqref{eq:rhs_nBE_diagonal} 

\be
\bGamma_{ij \rightarrow ka} \approx n_i n_j \langle \sigma_{ij \rightarrow ka} v_{\rm Mol}\rangle \, ,
\ee
Thus, similarly to the case of decays in Sec~\ref{apx:decay} in order to get from the fBE to nBE one has to assume that the distribution of axions has a thermal shape and that the quantum corrections can be neglected. 

In the simplest case when $f_i$ and $f_j$ have the MB shape the integrand in Eq.~\eqref{eq:sigmaV_general} only depends on $E_i$, $E_j$ and the angle between $\vec{p}_i$ and $\vec{p}_j$ (or $s$-invariant). Using the fact that $f_i \cdot f_j = \exp ((E_i+E_j)/T)$ one can change variables to $E_+ = E_i + E_j$ and $E_- = E_i - E_j$ and integrate analytically over them. The result is a \textit{1-dimensional} (1D) integration that has to be computed numerically

\be
\langle \sigma_{ij \rightarrow ka} v_{\rm Mol}\rangle = \frac{T}{8\pi^4n^{\rm eq}_i n^{\rm eq}_j} \int_{(m_i+m_j)^2}^{\infty} ds \sqrt{s} \; p^2_{ij}(s) \, K_1 \left( \sqrt{s}/T\right) \sigma_{ij \rightarrow ka}(s) \, ,
\label{eq:AnnRateGG}
\ee 
where $n^{\rm eq}_{i,j}$ are given by Eq.~\eqref{eq:density_MB} and $K_1$ is the modified Bessel functions of the first kind. 

Now let us get back to the $\gamma_{ij \rightarrow ak}$ function in Eq.~\eqref{eq:SmallGammaAnn}. In the most general case it can be reduced to a 4D integral. One possible way of reduction is to compute the integral over $d\Pi_i$ and $d\Pi_j$ in the CM frame and transforming the distribution functions via the following substitution 

\be
E_i = \frac{E^*_i + p^*_i v \cdot \cos \theta_{\rm CM}}{\sqrt{1 - v^2}} \, ,\quad E_j = \frac{E^*_j - p^*_i v \cdot \cos \theta_{\rm CM}}{\sqrt{1 - v^2}} \, ,
\label{eq:ei_ej_cm_frame}
\ee
where the asterisk denotes the quantities in the CM frame, $v = \sqrt{1 - s/(E_k+E_a)^2}$ is the velocity of the CM frame and $\theta_{\rm CM}$ is the angle between the momentum in CM frame and the velocity of the CM frame.  After the integration over $d\Pi^*_j$ and $E^*_i$ are performed with the use of the delta-function the residual angular integrals can be expressed via Mandelstam variable $t$ and an angle $\phi$ between the projections of $\vec{p}_a$ and $\vec{p_i^*}$ on the plane that is orthogonal to $\vec{p_a^*}$. The integration over $d\Pi_k$ has to be performed in the plasma frame reducing it to the integral over the energy $E_k$ in the plasma frame and $s$ variable. The cosine of $\theta_{\rm CM}$ can be expressed via the four integration variables that we are left with. Finally one arrives at the following expression

\begin{multline}
\gamma_{ij \rightarrow ak} = \frac{1}{p_a} \int^{\infty}_{E^{\rm min}_k} dE_k \; \frac{\lp 1 \pm f_k(E_k) \rp}{16 \, (2\pi)^4} \int^{s_{\rm max}}_{s_{\rm min}} \frac{ds}{p^*_k\sqrt{s}} \times \\ 
\times \int^{t_{\rm max}}_{t_{\rm min}}  dt \MM_{ij \rightarrow ak} \int^1_{-1} d\cos{\phi} \; \frac{f_i^*(E_i) f_j^*(E_j)}{\sqrt{1 - \cos{\phi}^2}} \, ,
\label{eq:gamma_2to2_gen_apx}
\end{multline}
where $E^{\rm min}_k$ is derived from the condition that $s \geq (m_i+m_j)^2$ and $p^*_k$ is the momentum of $k$ particle in the CM frame. The Mandelstam variables we use here can be expressed as follows 

\bea
s = (\mathcal{P}_a + \mathcal{P}_k)^2 &=& m^2_a + m^2_k + 2E_k E_a - 2p_a p_k \cos \theta_s \, ; \\
t = (\mathcal{P}_a - \mathcal{P}_i)^2 &=& m^2_i + m^2_k - 2E^*_i E^*_k + 2p^*_i p^*_k \cos \theta_t \, .
\eea
The energies in the distribution functions $f^*_i$ and $f^*_j$ are derived via the variables of integration in the following manner

\be 
E_{i,j} = \frac{E^*_{i,j}}{\sqrt{1 - v^2}} \pm \frac{p^*_{i,j} \left[ p^*_k \cos \theta_t - p_a \left( \cos\theta_t \cos \theta_{\rm aCM} - \sin\theta_t \sin\theta_{\rm aCM}\cos\phi \right) \right] }{E_k + E_a} \, ,
\ee
where $\theta_{\rm aCM}$ is the angle between $\vec{v}$ and $\vec{p_a}$ and is given by 

\be
\cos\theta_{\rm aCM} = \frac{ p_a^2 + \gamma^L (\vec{p_a} \cdot \vec{v}) (\gamma^L (\vec{p_a} \cdot \vec{v})/(\gamma^L+1) - E_a)}{p_a p^*_k} \, ,
\ee
with $\gamma^L = 1/\sqrt{1 - v^2}$ being the Lorentz factor and the scalar product $(\vec{p_a} \cdot \vec{v}) = (p_a^2 + p_a p_k \cos \theta_s)/(E_x+E_k)$. 

If we assume that $i$ and $j$ particles can be approximated by the MB statistics Eq.~\eqref{eq:gamma_2to2_gen_apx} drastically simplifies because the product of distributions can be transformed as follows 

\be
f_i \cdot f_j \approx \exp \left(-\frac{E_i + E_j}{T} \right)  = \exp \left(-\frac{E_a + E_k}{T} \right) \, .
\ee
and taken outside of the integration. Then the residual integrals can be expressed via the cross section in the following simple way 

\be
\gamma_{ij \rightarrow ak} = \frac{\exp(-q)}{(2\pi)^2} \int^{\infty}_{E^{\rm min}_k} dE_k \; E_k \, f_k(E_k) \int^{s_{\rm max}}_{s_{\rm min}} ds \; \sigma_{ak \rightarrow ij} (s) \, v_{\rm Mol} \, .
\label{eq:gamma_2to2_simp_apx}
\ee

\subsection{Annihilation}

The squared amplitude\footnote{The amplitudes for annihilation and Primakoff scattering were taken from Ref.~\cite{DEramo:2018vss}} for muon annihilation into axion and photon \textbf{summed} over initial and final state polarizations is given by 

\be
\MM_{\mu\mu \rightarrow a\gamma} = 16\pi \alpha \lp \frac{C_{\mu}}{f} \rp^2 \frac{m^2_{\mu}s^2}{(m^2_{\mu} - t)(s + t - m^2_{\mu})} \, ,
\ee
where $\alpha$ if the fine-structure constant. The cross section for this process is 

\be
\sigma_{\mu\mu \rightarrow a\gamma} = \alpha \lp \frac{C_{\mu}}{f} \rp^2 \; \frac{m^2_{\mu} \cdot {\rm artanh} \lp \sqrt{1 - \frac{4m^2_{\mu}}{s}} \rp }{s - 4m^2_{\mu}} \, .
\ee
The simplified formula for $\gamma$ from Eq.~\eqref{eq:gamma_2to2_simp_apx} yields the following 1D integral
\begin{multline}
\gamma_{\mu\mu \rightarrow a\gamma} (x,q) \approx \alpha \lp \frac{C_{\mu} m_{\mu}}{f} \rp^2 \frac{\exp(-q) \cdot T^2}{(2\pi)^2 q} \times \\
\times \int^{\infty}_{x^2/q} d\epsilon_k \; \frac{(2\epsilon_k q - x^2) \cdot {\rm artanh} \lp 1 - \frac{x^2}{\epsilon_k q}\rp - \epsilon_k q \sqrt{1 - \frac{x^2}{\epsilon_k q}} }{\exp(\epsilon_k) - 1} \, ,
\end{multline}
where $\epsilon_k = E_k/T$.

\subsection{Primakoff scattering} 

The squared amplitude for Primakoff scattering of axions on muons \textbf{summed} over initial and final state polarizations is given by 

\be
\MM_{\mu \gamma \rightarrow \mu a} = 16\pi \alpha \lp \frac{C_{\mu}}{f} \rp^2 \frac{m^2_{\mu}t^2}{(s - m^2_{\mu})(s + t - m^2_{\mu})} \, .
\ee
The cross section for this process is 

\be
\sigma_{\mu \gamma \rightarrow \mu a} = \alpha \lp \frac{C_{\mu}}{f} \rp^2 \; \frac{m^2_{\mu} \lp 2s^2 \log (s/m^2_{\mu}) - 3s^2 + 4m^2_{\mu} s - m^4_{\mu}\rp }{8s^2(s - m^2_{\mu})} \, .
\ee

\begin{multline}
\gamma_{\mu\gamma \rightarrow \mu a} (x,q) \approx \alpha \lp \frac{C_{\mu} m_{\mu}}{f} \rp^2 \frac{\exp(-q) \cdot T^2}{4 (2\pi)^2 q} \times \\
\times \left. \int^{\infty}_{x} d\epsilon_k \; \frac{2 \tilde{s} \log(\tilde{s}/x^2) + 4x^2 \log(\tilde{s}) - 5\tilde{s} + x^4/\tilde{s}}{\exp(\epsilon_k) + 1} \right|^{\tilde{s}_{\rm max}}_{\tilde{s}_{\rm min}} \, ,
\end{multline}
where $\tilde{s} = s/T^2$ and $\tilde{s}_{\rm min/max} = x^2 + 2\epsilon_k q \pm 2 \sqrt{\epsilon_k^2 - x^2}q$.

\bibliographystyle{JHEP}
\bibliography{References}

\end{document}